\documentclass[10pt,prl,aps,twocolumn,floatfix]{revtex4}
\usepackage{graphics}
\usepackage{amssymb}
\usepackage{color}
\usepackage{amsmath,hyperref}
\usepackage{amsbsy}
\usepackage{amsmath}
\usepackage{epsfig}
\newcommand\bea{\begin{eqnarray}}
\newcommand\eea{\end{eqnarray}}
\newcommand\beq{\begin{equation}}
\newcommand\eeq{\end{equation}}

\newcommand{\non}{\nonumber}

\newcommand{\al}{\alpha}
\newcommand{\be}{\beta}
\newcommand{\de}{\delta}
\newcommand{\De}{\Delta}
\newcommand{\ga}{\gamma}
\newcommand{\lam}{\lambda}
\newcommand{\si}{\sigma}
\newcommand{\ta}{\theta}
\newcommand{\da}{\dagger}
\newcommand{\ha}{{\hat a}}
\newcommand{\hb}{{\hat b}}
\newcommand{\he}{{\hat e}}

\newcommand{\vs}{{\vec \sigma}}

\begin{document}

\title{Aperiodically driven integrable systems and their emergent steady 
states}

\author{Sourav Nandy$^1$, Arnab Sen$^1$, and Diptiman Sen$^2$}

\affiliation{$^1$Department of Theoretical Physics, Indian Association
for the Cultivation of Science, Jadavpur, Kolkata 700032, India \\
$^2$Centre for High Energy Physics, Indian Institute of Science, Bengaluru 
560012, India}

\begin{abstract}
Does a closed quantum many-body system that is continually
driven with a time-dependent Hamiltonian finally reach a steady state? 
This question has only recently been answered for driving protocols that 
are periodic in time, where the long time behavior of the 
local properties synchronize with the drive and can be described by an 
appropriate periodic ensemble. Here, we explore the consequences of breaking 
the time-periodic structure of the drive with additional aperiodic noise in a 
class of integrable systems. We show that the resulting unitary dynamics leads 
to new emergent steady states in at least two cases. While any typical 
realization of random noise causes eventual heating to an infinite temperature 
ensemble for all local properties in spite of the system being integrable, 
noise which is self-similar in time leads to an entirely different steady 
state, which we dub as ``geometric generalized Gibbs ensemble'', that emerges 
only after an astronomically large time scale. To understand the approach 
to steady state, we study the temporal behavior of certain coarse-grained 
quantities in momentum space that fully determine the reduced density matrix 
for a subsystem with size much smaller than the total system. Such quantities 
provide a concise description for any drive protocol in integrable systems 
that are reducible to a free fermion representation. 
\end{abstract}

\date{\today}

\maketitle

\section{Introduction and Motivation}
\label{secI}

\begin{figure}
{\includegraphics[width=\hsize]{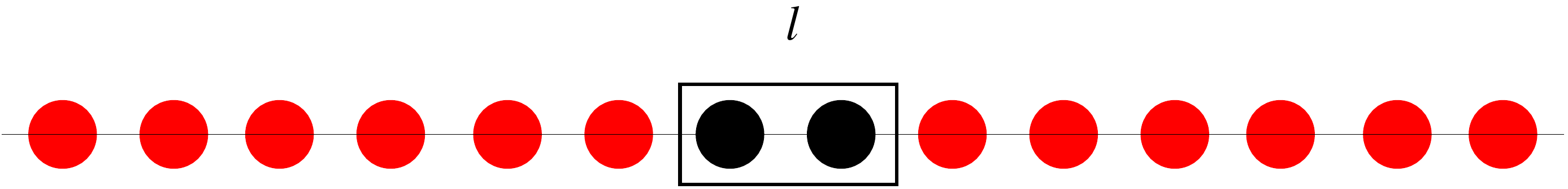}} 
\caption{The separation of a closed system into a subsystem with $l$ sites 
(indicated in black) and the environment with $L-l$ sites (indicated in red). 
Local properties only involve degrees of freedom on the sites of a subsystem 
that satisfies $l \ll L$ with $L \gg 1$.} \label{fig0main} \end{figure}

Statistical mechanics can be derived from Jaynes' principle of maximum 
entropy~\cite{Jaynes1957a, Jaynes1957b} given the constants of motion for any 
generic many-body system. Remarkably, this principle may also be extended to 
the level of a single eigenstate in a thermodynamically large, isolated 
quantum system leading to the eigenstate thermalization hypothesis 
~\cite{Deutsch1991, Srednicki1994, RigolDO2008}. The eigenstate thermalization 
hypothesis implies that the expectation values of all operators which are 
local in space equal those obtained from a maximum entropy statistical 
description when {\it appropriate} conserved quantities are taken into account 
for constructing the ensemble. The Hamiltonian is the sole such conserved 
quantity in a generic chaotic system. Exceptions to the eigenstate 
thermalization hypothesis are provided by disordered systems that show 
localization, both for non-interacting models~\cite{Anderson1958} 
and with interactions~\cite{BaskoAA2006} (i.e., many-body localized 
systems~\cite{NandkishoreH2015}). Integrable models provide examples of 
systems that are not localized and yet do not satisfy the eigenstate 
thermalization hypothesis. However, Jaynes' principle may again 
be applied here by considering the extensive number of conservation laws 
due to integrability that then leads to a generalized
Gibbs ensemble (GGE)~\cite{RigolDYO2007, CassidyCR2011, CauxE2013} 
instead of the standard Gibbs ensemble. These statistical descriptions are 
expected to hold only for local properties (i.e., properties that are 
determined by the degrees of freedom in a subsystem that is much smaller than 
the rest of the system, as shown schematically in Fig.~\ref{fig0main}), and 
not for the wave function of the entire system which is 
after all in a pure state~\cite{PopescuSW2006}.

Tremendous experimental progress in producing and manipulating well-isolated 
quantum systems such as ultracold quantum 
gases~\cite{Blochreview2005, Goldman2014, Langenetalreview2015, Eck17}, 
has also led to great interest in driven closed quantum systems. Removing 
the time translational invariance of closed quantum systems via a 
time-dependent Hamiltonian leads to the much richer possibility of stabilizing 
non-equilibrium steady states with purely unitary dynamics. 
Such non-equilibrium steady states may even have properties otherwise 
forbidden in thermal equilibrium,
such as spontaneous time-translational symmetry breaking in many body localized
systems~\cite{ElseBN2016, KhemaniLMS2016} and dynamical topological 
ordering~\cite{KitagawaBRD2010, NathanR2015}. 
Many-body Floquet systems where the time-dependent Hamiltonian satisfies
\bea H(t)=H(t+nT), \label{periodicdrive1} \eea
with $T$ being the time period of the drive and $n$ being integer-valued, 
is a well-studied case where such non-equilibrium steady states have been 
shown to exist. 
More precisely, the local properties of Floquet systems eventually synchronize 
with the period of the drive that then allows the possibility of a ``periodic''
ensemble~\cite{LazaridesDM2014a, LazaridesDM2014b, Ponte15b, AlessioR2014}. 
Propagating the wave function $|\psi \rangle$ of the system in a stroboscopic 
manner
\bea |\psi(nT) \rangle = U(T)^n |\psi(0) \rangle, \label{periodicdrive2} \eea
where $U(T)$ is the time evolution operator for one time period, generates 
a discrete quantum map indexed by $n$, and the steady state emerges as 
$n \to \infty$. This steady state can again be described using 
Jaynes' principle~\cite{LazaridesDM2014a} where only 
quantities that are stroboscopically conserved need to be taken into account. 
In the non-integrable case, it is expected that no such conserved quantity 
exists~\cite{NandkishoreH2015} since the Hamiltonian is no longer conserved; 
hence the system {\it should} locally mimic an infinite temperature ensemble 
in the steady state by entropy maximization. This is indeed consistent 
with the results of several recent studies~\cite{LazaridesDM2014b, Ponte15b, 
AlessioR2014, Russomanno12, Bukov15, Ponte15a, Eckardt15}, but a complete 
understanding is still lacking. The situation is markedly different for 
integrable systems. For systems that are reducible to free fermions e.g., 
the prototypical one-dimensional (1D) spin-1/2 transverse field Ising model, 
it has been shown~\cite{LazaridesDM2014a} that an extensive number of 
quantities that are stroboscopically conserved continue to exist {\it even} 
in the driven case; these prevent the system from approaching the infinite 
temperature ensemble locally and instead lead to a {\it periodic} GGE 
(p-GGE) description of the synchronized local properties at late times. The 
fate of other types of integrable systems like Bethe-integrable ones (for 
instance, the 1D spin-1/2 $XXZ$ model) under periodic drives is still an open 
issue~\cite{BurkovH2012}.

\subsection{Synopsis}
\label{subsecIa} 

In this work, we consider integrable models that are reducible to free 
fermions, and our objective is to understand whether well-defined 
non-equilibrium steady states exist for 
continually driven protocols that are {\it not} periodic functions in time. We 
restrict ourselves to drives that can be represented by discrete quantum maps 
similar to Eq.~\eqref{periodicdrive2} for ease of analysis. For concreteness, 
we work here with the 1D transverse field Ising model which is a standard 
many-body integrable quantum system~\cite{Subirbook,SIBKCbook,Duttabook}. The 
time-dependent Hamiltonian is given by
\bea H(t) ~=~ - ~\sum_{j=1}^L ~\left[ g(t) s_j^x+ s_j^z s_{j+1}^z \right],
\label{modeldef} \eea 
where $s^{x,y,z}$ are the usual Pauli operators, $L$ denotes the number of 
spins in the system under consideration, and we assume periodic boundary 
conditions in space. $g(t)$ represents the time-dependent
transverse magnetic field through which the system is continually driven.

We start with a reference function $g_{\mathrm{ref}}(t)$ defined for 
$0 \le t \le T$. Repeatedly using the corresponding time evolution operator 
$U(T)$ for all values of $n>0$ in Eq.~\eqref{periodicdrive2} results in 
periodic driving with period $T$. Consider now a different drive protocol 
where $g(t)$ is constructed by patching together rescaled versions of 
$g_{\mathrm{ref}}(t)$ in time as a function of $n$ such that 
the period of $g_{\mathrm{ref}}(t)$ is stretched or reduced by $dT$ 
depending on $n$. The corresponding time evolution operator is 
then denoted by $U(T+dT)$ or $U(T-dT)$. The new drive protocol $g(t)$ can now 
be represented by a sequence in $n$ that takes a value of either $+1$ 
(representing $U(T+dT)$) or $-1$ (representing $U(T-dT)$) at each $n$. 
There are an {\it infinite} number of ways to choose the sequence of $\pm 1$ 
as a function of $n$, each of which represents a different purely unitary 
dynamics, such that the resulting driving protocol $g(t)$ cannot be 
represented by {\it any} periodic function in time.
We note that the protocol described above involving two time periods
$T+dT$ and $T-dT$ is mathematically equivalent to a protocol in which
the time period $T$ is kept fixed but the Hamiltonian is scaled globally
by factors of $1+dT/T$ and $1-dT/T$ respectively. The important point about
the protocol is that it involves two possible time evolution operators which
do not commute with each other.

We will work out the nature of the resulting non-equilibrium steady states as 
$n \to \infty$ 
in two cases here. First, we will consider a typical realization of a random 
sequence of $\pm 1$ where the sign is chosen with probability 1/2 at each 
$n$ with $dT \ll T$. This mimics a periodic drive protocol perturbed with 
random noise due to the lack of perfect control over the time variation of 
$g(t)$ in an experimental set-up. Secondly, we will consider a different 
sequence of $\pm 1$ that is neither random nor periodic in $n$ but has 
a fractal structure instead, and thus can be 
thought of as an example of scale-invariant noise. 

For a periodic drive protocol perturbed with random noise where $dT \ll T$, 
we will show that the system, after initially approaching the p-GGE that 
corresponds to $dT=0$, eventually heats up to an infinite temperature ensemble 
as $n \to \infty$ in spite of the underlying integrability of the model. The 
time spent by the system close to the p-GGE, which can now be thought of as 
a quasisteady {\it prethermal} state, and the eventual manner of heating up to 
an infinite temperature ensemble can both be understood from a simple random 
walk argument, provided that $dT \ll T$. We will
further show that a {\it coarse-graining procedure} in momentum space 
explains how the reduced density matrix of any subsystem with $l$ adjacent 
spins approaches an infinite temperature ensemble as long as $l \ll L$ 
even though the dynamics is completely unitary. 
It is well known that weakly interacting systems (i.e., systems close 
to an integrable point) after a sudden perturbation show a prethermalization 
regime that can be understood by a GGE based on perturbatively constructed 
constants of motion~\cite{KollarWE2011}, before eventually thermalizing on 
a much longer time scale~\cite{MoeckelK2008, Gring2012} with the energy being 
the only constant of motion. In the perturbed Floquet integrable 
systems with random noise that we consider, the integrability of the 
model is never broken but the stroboscopically conserved quantities 
at $dT=0$ no longer remain conserved when $dT \neq 0$.

However, it is important to note that the lack of conservation of these 
quantities does {\it not} fix the form of the final non-equilibrium steady 
state for an aperiodic 
drive protocol. For a periodic drive protocol perturbed with scale-invariant 
noise, we will show that if the local properties are observed not as a function
of $n$, but instead as a function of $2^n$ (thus, geometrically instead of 
stroboscopically), a new steady state emerges when $n \to \infty$ which we call
``geometric'' GGE (g-GGE). Interestingly, this steady state can also be 
understood using an effective p-GGE construction even though the drive in not 
periodic in time because of a fixed point property of products of $2 \times 2$
$SU(2)$ matrices. An extensive number of conserved quantities appear when the 
system is viewed after every $2^n$ drives, but only when $n$ is sufficiently 
large, which shows that these conservations are {\it emergent}.

The rest of the paper is arranged in the following manner. In Sec.~\ref{secII},
we review some results relevant for our work and set the notations for the 
rest of the paper. Sec.~\ref{subsecIIa} details the pseudospin representation 
for the 1D transverse field Ising model that allows the wave function of the 
system to be denoted in terms of points on Bloch spheres in momentum space. 
Sec.~\ref{subsecIIb} details the aperiodic driving protocols that we study. 
Sec.~\ref{subsecIIc} reviews the way to obtain the reduced density matrix of a 
subsystem of $l$ adjacent spins using appropriate correlation matrices. In 
Sec.~\ref{secIII} (including Sec.~\ref{subsecIIIa} and Sec.~\ref{subsecIIIb}),
we introduce certain coarse-grained quantities in momentum space which fully 
determine the reduced density matrix of 
a subsystem when $l \ll L$ and thus {\it all} local quantities, and we discuss
their behavior for time-periodic drive protocols. In Sec.~\ref{secIV}, we 
present results for perturbed Floquet systems with random noise and use a 
random walk argument to show how the system heats up to an infinite temperature ensemble in spite of its
integrability. In Sec.~\ref{secV}, we consider perturbed Floquet systems with 
noise which is scale-invariant in time and show that for a certain type of 
scale-invariant noise, the system reaches a steady state which is 
{\it not} an infinite temperature ensemble, but it does so only 
after an astronomically large number of drives. The final steady state is 
explicitly constructed using a fixed point property of strings of $2\times 2$ 
$SU(2)$ matrices. Finally, we summarize our results and conclude with some 
future directions in Sec.~\ref{secVI}.

\section{Preliminaries}
\label{secII}

\subsection{Pseudospin representation of dynamics for the 1D transverse 
field Ising model}
\label{subsecIIa}

To solve the 1D transverse field Ising model (Eq.~\eqref{modeldef}), we 
introduce the standard Jordan-Wigner transformation of spin-1/2's to spinless 
fermions~\cite{Lieb61}
\bea s_n^x &=& 1-2c^\da_n c_n, \non \\ 
s_n^z &=& -(c_n+c_n^\da)\prod_{m<n}(1-2c^\da_m c_m). \label{jw1} \eea
Writing $H$ in Eq.~\eqref{modeldef} in terms of these fermions, we obtain
\bea H ~=~ P^+ H^+ P^+ ~+~ P^-H^-P^-, \label{jw2} \eea
where $P^{\pm}$ are projectors on to subspaces with even ($+$) 
and odd ($-$) number of fermions, and 
\bea H^{\pm} &=& -\sum_{n=1}^L ~[g(t) ~-~ 2g(t) c^\da_n c_n ~+~ 
c^\da_n c_{n+1} ~+~ c_{n+1}^\da c_n \non \\
&& ~~~~~~~~~+ ~c_{n+1}c_n ~+~ c_n^\da c_{n+1}^\da], \label{jw3} \eea
with antiperiodic (periodic) boundary conditions for the fermions for 
$H^+$ ($H^-$). We now focus on the case of even $L$ with antiperiodic 
boundary conditions for the fermions (which corresponds to 
periodic boundary conditions for the spins). To go to $k$ space, we define
\bea c_k &=& \frac{e^{i\pi/4}}{\sqrt{L}} ~\sum_{n=1}^L ~e^{-ikn} ~c_n, \non \\
c_n &=& \frac{e^{-i\pi/4}}{\sqrt{L}} ~\sum_k ~e^{ikn} ~c_k, \label{jw4} \eea 
where $k = 2\pi m/L$ with $m = -(L-1)/2, \cdots, -1/2$, $1/2, \cdots, (L-1)/2$ 
for even $L$. Rewriting $H^+$ in terms of $c_k$ and $c_k^\da$, we get
\bea H^+ &=& \sum_k ~[2(g(t)-\cos k) ~c^\da_k c_k ~+~\sin k~ c_{-k}c_k \non \\
&& ~~~~~~+ ~\sin k ~c^\da_k c^\da_{-k}-g(t)]. \label{fermionH} \eea
This Hamiltonian connects the vacuum $|0 \rangle$ of the $c$ fermions with the
two-particle state $|k,-k \rangle =c_k^\da c^\da_{-k}|0 \rangle$, and 
$|k \rangle 
= c_k^\da |0 \rangle$ with $|-k \rangle = c_{-k}^\da |0 \rangle$. It is 
more convenient to write Eq.~\eqref{fermionH} as $H = \sum_{k>0} H_k$ where
\bea H_k &=& 2(g(t)-\cos k) [c^\da_k c_k - c_{-k}c^\da_{-k}] \non \\
&+& 2 \sin k [c_{-k}c_k + c^\da_k c^\da_{-k}]. \label{pseudospin1} \eea

We now introduce a ``pseudospin representation'' $\si_k$ (which is different 
from the original $s$ matrices in Eq.~\eqref{modeldef}) 
where $|\uparrow \rangle_k = |k,-k \rangle = c_k^\da c^\da_{-k}|0 
\rangle$ and $|\downarrow \rangle_k = |0 \rangle$~\cite{KolodrubetzCH2012}. 
Writing $H_k$ in Eq.~\eqref{pseudospin1} in this basis, we see that
\bea H_k ~=~ 2(g(t)-\cos k) ~\si^z_k ~+~ 2 \sin k ~\si^x_k. 
\label{pseudospin2} \eea 
Then the pseudospin state $|\psi_k \rangle$ for each $k$ mode evolves
independently as
\bea i \frac{\mathrm{d}}{\mathrm{dt}} |\psi_k \rangle ~=~ H_k(t)
|\psi_k \rangle, \label{pseudospin3} \eea
where
\bea |\psi(t) \rangle &=& \otimes_{k>0}|\psi_k(t) \rangle, \non \\
|\psi_k (t)\rangle &=& u_k(t)|\uparrow \rangle_k + v_k(t)|\downarrow \rangle_k.
\label{pseudospin4} \eea
Thus, specifying $(u_k(t),v_k(t))^T$ (where the superscript $T$ denotes the
transpose of the row $(u_k(t),v_k(t))$) for each allowed $k>0$ for a finite 
$L$ specifies the complete wave function $|\psi(t) \rangle$ through 
Eq.~\eqref{pseudospin4}. The wave function of the system can be equivalently 
characterized by a two-dimensional state space at each allowed positive $k$, 
and $(u_k(t),v_k(t))^T$ can then be represented as a point that evolves with 
time according to Eq.~\eqref{pseudospin3} on the corresponding Bloch sphere.

In the rest of the paper, we will take $(u_k,v_k)^T=(0,1)^T$ to be the starting 
state at each $k$ from which the unitary dynamics proceeds. This corresponds 
to the system being initially prepared in the pure state where all the spins 
have $\si^x=+1$ (which corresponds to the ground state for $g \to \infty$).

\subsection{Details of the driving protocol}
\label{subsecIIb}

While our basic conclusions are protocol independent and do not depend 
on the exact nature of $g_{\mathrm{ref}}(t)$, we take it to 
have a square pulse variation in time for mathematical convenience, namely,
\bea g_{\mathrm{ref}} &=& g_i \mbox{~~~for~~~} 0 \le t<T/2 \non \\
&=& g_f \mbox{~~~for~~~} T/2 \le t \le T. \label{squarepulse} \eea
The corresponding unitary time evolution operator $U_k(T)$ 
for the pseudospin mode with momentum $k$ equals 
\bea U_k(T) = \exp \left(-iH_k(g_f)\frac{T}{2}\right) \exp \left(-iH_k(g_i)
\frac{T}{2} \right), \label{defU} \eea
where $H_k$ has the form given in Eq.~\eqref{pseudospin2}. For the perturbed 
Floquet dynamics that we consider in this work, we need to define 
$U_k(T \pm dT)$; these follow in a straightforward manner from Eq.~\eqref{defU}
by replacing $T \to T \pm dT$.

Rewriting Eq.~\eqref{periodicdrive2} for the perturbed Floquet dynamics for 
the mode with momentum $k$, we get 
\begin{widetext}
\bea |\psi_k(n) \rangle ~=~ U_k(T+ \tau_n dT) U_k(T + \tau_{n-1}dT) \cdots 
U(T + \tau_1 dT) |\psi_k(0) \rangle ~=~ {\cal T} ~\prod_{i=1}^n ~U_k(T+\tau_i
dT) |\psi_k(0) \rangle, \label{defsequence} \eea
\end{widetext}
where the sequence ${\tau_i}=\tau_1, \tau_2, \tau_3, \cdots$ (with each 
$\tau_i$ being equal to either $+1$ or $-1$) is the same for all the allowed 
$k$ modes at a finite $L$. (In the last expression in Eq.~\eqref{defsequence}, 
${\cal T}$ denotes time-ordering).

For Floquet systems perturbed with random noise, the sequence 
${\tau_i}$ can be taken to be any typical realization of a random 
process where each element is chosen to be $\pm 1$ randomly and 
independently with probability 1/2; we will discuss the results 
for such a realization later in the paper. For Floquet systems perturbed with 
scale-invariant noise, we take the well-known Thue-Morse sequence
~\cite{Thue1906, Morse1921a, Morse1921b} which is an infinite sequence of 
$\tau_i = \pm 1$ that is obtained by starting with $\tau_1=-1$ and 
successively appending the negative of the sequence obtained thus far. 
The first few steps of this recursive procedure yield
\begin{widetext}
\bea 
\label{TMS}
m=0:&& \mbox{~~} \tau_1 ~=~ -1 \\ \non
m=1:&& \mbox{~~} \tau_1,\tau_2 ~=~ -1,+1 \\ \non
m=2:&& \mbox{~~} \tau_1, \cdots, \tau_4 ~=~ -1,+1,+1,-1 \\ \non 
m=3:&& \mbox{~~} \tau_1, \cdots, \tau_8 ~=~ -1,+1,+1,-1,+1,-1,-1,
+1 \\ \non 
m=4:&& \mbox{~~} \tau_1, \cdots, \tau_{16} ~=~ -1,+1,+1,-1,+1,-1,-1,+1,+1,-1,
-1,+1,-1,+1,+1,-1 \\ \non 
\vdots 
\eea
\end{widetext}
At each recursion level $m$, we thus obtain the first $2^m$ elements of this 
infinite sequence. The self-similar structure of the Thue-Morse sequence can be
verified by removing every second term from it, which then results in the same 
infinite sequence. This sequence is thus an example of a {\it quasiperiodic} 
sequence in time which is neither periodic nor random. Note that the average 
time period of this perturbed Floquet system equals $T$ (the time period 
of the unperturbed Floquet system), just like in the random noise case.

\subsection{Reduced density matrices and the distance measure}
\label{subsecIIc}

Given the wave function $|\psi(n) \rangle$ of the entire system, which can be 
described by the corresponding density matrix $\rho = |\psi(n) \rangle \langle 
\psi(n)|$, all local properties within a length scale $l$ can be understood by 
considering the reduced density matrix of a subsystem of $l$ adjacent 
spins (Fig.~\ref{fig0main}); we denote this by $\rho_l$ which is given by
\bea \rho_l ~=~ \mathrm{Tr}_{L-l} (\rho), 
\label{RDM} \eea
where all the other $L-l$ degrees of freedom have been integrated out. Even 
though $\rho$ is a pure density matrix at any $n$, the reduced density matrix 
$\rho_l$ is mixed since the pure state $| \psi(n) \rangle$ typically gets 
more entangled as $n$ increases.

The reduced density matrix of $l$ adjacent spins when $|\psi(n) \rangle$ has 
the form given in Eq.~\eqref{pseudospin4} is determined in terms of the 
correlation functions of the $c$ fermions at these $l$ 
sites~\cite{VidalLRK2003}. For free fermions, since all higher-point 
correlation functions are expressible using two-point correlations from 
Wick's theorem, the reduced density matrix can be determined from the 
knowledge of two $l \times l$ matrices~\cite{Peschel2003}, $\mathbf{C}$ and 
$\mathbf{F}$, whose elements are constructed from $u_k(n)$ and $v_k(n)$ as 
follows:
\bea C_{ij} &=& \langle c^\da_i c_j \rangle_n = \frac{2}{L} \sum_{k>0} 
|u_k(n)|^2 \cos[k (i- j)], \label{matrices1} \non \\
F_{ij} &=& \langle c^\da_i c^\da_j \rangle_n = \frac{2}{L} 
\sum_{k >0} u^*_k(n) v_k(n) \sin[k(i-j)], \label{corrmat1} \eea 
where $i,j$ refer to sites in the subsystem. Using these expressions, 
we construct the $2l \times 2l$ matrix ${\mathcal C}_n(l)$ given by
\bea {\mathcal C}_n(l) &=& \left( \begin{array}{cc} \mathbf{I-C} &
\mathbf{F} \\ \mathbf{F}^{\ast} & \mathbf{C } \end{array} \right).
\label{corrmat2} \eea
The reduced density matrix $\rho_l$ is then completely fixed by the 
eigenvalues and eigenvectors of $\mathcal{C}_n(l)$. For 
example, the entanglement entropy $S_{\mathrm{ent}}(l)$ equals:
\bea S_{\mathrm{ent}}(l)&=& -~ \mathrm{Tr} [\rho_l \log(\rho_l)] \non \\
&=& -~ \sum_{k=1}^{2l} ~p_k \log(p_k) \label{Sent} \eea
where $p_k$ denotes the $k$-th eigenvalue of $\mathcal{C}_n(l)$.

If a non-equilibrium steady state exists as 
$n \to \infty$, then the convergence of the 
local properties to their final steady state values as $n$ increases can be 
characterized by defining a distance measure $\mathcal{D}_n(l)$ as
\bea \mathcal{D}_n(l) ~=~ \frac{1}{2l}\mathrm{Tr}\sqrt{[\mathcal{C}_\infty(l)-
\mathcal{C}_n(l)]^\da [\mathcal{C}_\infty(l)-\mathcal{C}_n(l)]}.
\label{distancemeasure} \eea
We note that $0 \leq \mathcal{D}_n(l) \leq 1$ and is identically zero only 
if $\mathcal{C}_n(l) = \mathcal{C}_\infty(l)$. For a periodically driven 
protocol with $dT=0$, $\mathcal{C}_\infty(l)$ can be calculated using the 
appropriate p-GGE; the distance measure decays to zero as $\mathcal{D}_n(l)
\sim n^{-\al (T)}$ for large $n$ (and $l \ll L$), 
with $\al(T)=1/2$ or $3/2$ depending on the time period $T$
of the drive~\cite{SenNS2016}. We note that $\mathcal{D}_n(l)$ does not 
decay to zero as $n \to \infty$ if $l/L$ is finite; this explicitly 
shows that only local quantities can be described by the p-GGE.

\section{Coarse-graining in momentum space}
\label{secIII}

Even though the entire wave function $|\psi(n) \rangle$ needs the 
specification of $(u_k(n),v_k(n))^T$ at $k=2\pi m/L$ with 
$m=1/2,3/2,\cdots, (L-1)/2$, $\mathcal{C}_n(l)$ and consequently the reduced 
density matrix for any $l \ll L$ depends {\it only} on certain 
{\it coarse-grained variables} defined as follows~\cite{LaiY2015, NandySDD2016}
\bea \left(|u_k(n)|^2\right)_c &=& \frac{1}{N_c}\sum_{k \in k_{\mathrm{cell}}} 
|u_k(n)|^2, \non \\
\left(u^*_k(n)v_k(n)\right)_c &=& \frac{1}{N_c}\sum_{k \in k_{\mathrm{cell}}} 
u^*_k(n) v_k(n). \label{coarsegrain1} \eea
These variables are defined using $N_c (\gg 1)$ consecutive $k$ modes that lie 
within a cell ($k_{\mathrm{cell}}$) which has an average momentum $k_c$ in
momentum space and a size $\de k$ where $1/L \ll \de k \ll 1/l$. With this 
condition on $\de k$, the sinusoidal factors in Eq.~\eqref{corrmat1} can be 
replaced to a very good approximation by 
\bea \cos[k(i-j)] &\simeq& \cos[k_c(i-j)], \mbox{~~} \forall \mbox{~} 0 
\leq |i-j| \leq l, \non \\
\sin[k(i-j)] &\simeq& \sin[k_c(i-j)], \mbox{~~} \forall \mbox{~} 0 \leq |i-j| 
\leq l, \eea
for all the $k$ modes within the cell. 
The sum over momentum in Eq.~\eqref{corrmat1} can thus be carried out in two 
steps, first summing over the consecutive momenta in a single cell, and then 
summing over the different momentum cells. This gives
\bea C_{ij} &\simeq& \frac{1}{\mathcal{N}_{\mathrm{cell}}} ~\sum_{k_c} 
~(|u_k(n)|^2)_c \cos [k_c(i-j)], \non \\
F_{ij} &\simeq& \frac{1}{\mathcal{N}_{\mathrm{cell}}} ~\sum_{k_c} 
~(u^*_k(n)v_k(n))_c \sin [k_c(i-j)], \label{coarsegrain2} \eea 
where $\mathcal{N}_{\mathrm{cell}} \ll L/2$ represents 
the number of momentum cells after the coarse-graining procedure.

Note that even though quantities like $|u_k(n)|^2$ and $u^*_n(k)v_n(k)$ 
continue to display Rabi oscillations and thus have no well-defined $n \to 
\infty$ limit, the coarse-grained variables $(|u_k(n)|^2)_c$ and $(u^*_k(n)
v_k(n))_c$ must reach final steady state values for a non-equilibrium steady 
state to exist. 

\begin{figure}
{\includegraphics[width=\hsize]{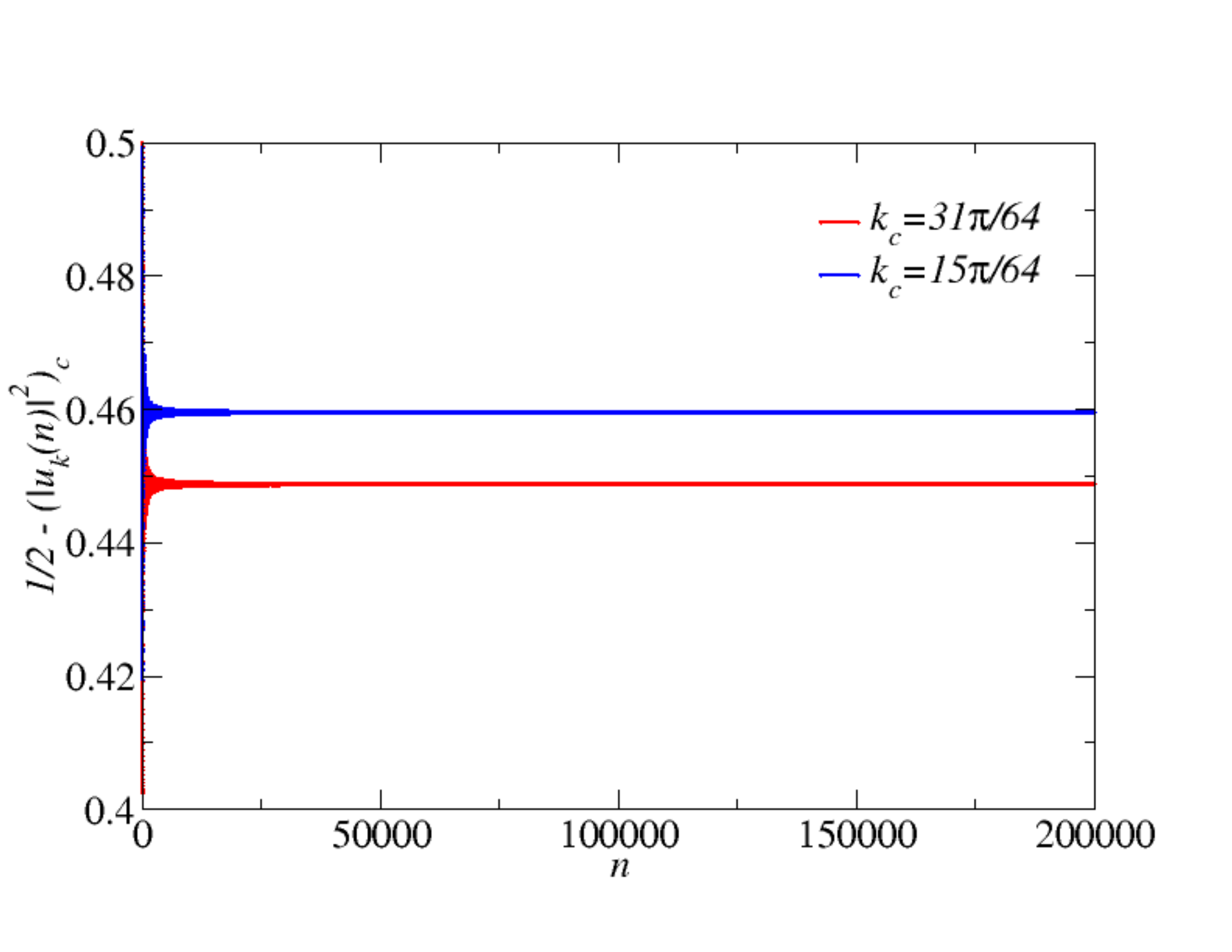}} 
\caption{Behavior of the coarse-grained quantities $(|u_k(n)|^2)_c$ as 
a function of $n$ for a perfectly periodic drive ($dT=0$). The parameters 
used are $L=131072$, $g_i=4$, $g_f=2$, and $T=0.2$. Each coarse-grained cell
contains $N_c = 2048$ consecutive momenta, and $k_c$ denotes the average 
momentum of such a cell. Results for $k_c=31\pi/64$ (lower line in red) and 
$k_c=15\pi/64$ (upper line in blue) are shown.} \label{fig1main} \end{figure}

If the system is locally described by an infinite temperature ensemble, then 
$\rho_l =I$, where $I$ is 
the identity matrix, for any $l \ll L$. This immediately implies that 
$(|u_k(n)|^2)_c \to 1/2$ (also $(|v_k(n)|^2)_c \to 1/2$) and $(u^*_n(k)v_n
(k))_c \to 0$ for all the coarse-grained momentum cells as $n \to \infty$, 
since $C_{ij}=(1/2) \de_{i,j}$ and $F_{ij}=0$ for $\rho_l=I$. (A similar 
behavior has been found for repeated quenching of the transverse field in the
1D spin-1/2 $XY$ model~\cite{Mukherjee08}). We calculate the $n$ dependence of 
such coarse-grained quantities numerically by taking a sufficiently large 
system size $L$ (typically $L=131072$, but $L=524288$ in some cases), 
so that the allowed momentum modes are sufficiently dense in the 
interval $k \in [0,\pi]$, and then dividing the Brillouin zone into 
$\mathcal{N}_{\mathrm{cell}} =32$ cells, each of which has $N_c \gg 1$ 
consecutive momenta. For the periodic drive case where $dT=0$, we see that 
although $(|u_k(n)|^2)_c$ indeed converges to a finite value 
(Fig.~\ref{fig1main}) and thus has a well-defined non-equilibrium steady state,
this constant depends on the value of the average momentum of the 
coarse-grained cell ($k_c$) and is different from 1/2, showing that the local 
description is {\it different} from that of an infinite temperature ensemble.

\subsection{Probability distribution on the Bloch sphere}
\label{subsecIIIa}

At each $k$, the pseudospin state $|\psi_k (n)\rangle$ obtained after $n$ 
drives can be parameterized in terms of three angles $(\ta_k(n),\phi_k(n),
\be_k(n))$,
\beq \left( \begin{array}{c}
u_k(n) \\
v_k(n) \end{array} \right) ~=~ e^{i\be_k(n)} ~\left( \begin{array}{c}
\cos (\ta_k(n)/2) e^{-i\phi_k(n) /2} \\
\sin (\ta_k(n)/2) e^{i\phi_k(n) /2} \end{array} \right). 
\label{blochsphere1} \eeq
The overall phase $\be_k(n)$ is not important and we will ignore it 
henceforth. We will concentrate on the angles $\ta_k(n)$ and $\phi_k(n)$ 
which define the motion of a point on the unit Bloch sphere for each momentum 
$k$. From Eq.~\eqref{defU}, it is clear that $U_k(T)$ is an element of the 
group $SU(2)$ (namely, $2 \times 2$ unitary matrices
with determinant equal to 1), and it can be written as
\beq U_k(T) ~=~ \exp (-i \ga_k \he_k \cdot \vs), \label{uk2} \eeq
where $0 \le \ga_k \le \pi$ and $\he_k = (e_{1k}, e_{2k}, e_{3k})$ is a unit 
vector. For a drive protocol that is perfectly periodic, $(\ta_k(n),\phi_k(n))$
traces out a circle on the Bloch sphere formed by the intersection of this 
unit sphere with the plane that contains the point $(\ta_k(0),\phi_k(0))$ and 
whose normal vector equals $\he_k$. It is then clear that 
the quantities $\mathcal{I}_k$ defined as
\bea \mathcal{I}_k ~=~ \langle \psi_k(n)| \he_k \cdot \vs | \psi_k(n) \rangle
\label{conservedIk} \eea
are all independent of $n$ for a periodic drive protocol and thus define 
the extensive number of stroboscopically conserved quantities.

The behaviors of the coarse-grained quantities in Eq.~\eqref{coarsegrain1} 
depend on the simultaneous positions of $(\ta_k(n), \phi_k(n))$ on the unit
sphere of all the momentum modes that lie within a coarse-grained cell. 
Since their number $N_c \gg 1$ when $L \gg 1$, it is useful to instead 
look at the probability distribution of these $N_c$ points on the unit sphere.
In particular, we will study the probability distribution of 
$|u_k(n)|^2-|v_k(n)|^2 = \cos \ta_k$ for each such coarse-grained momentum 
cell, which we denote as $P_n(\cos \ta_k)$, as a function of $n$.
We can also define another probability distribution from the motion 
of a single $k$ mode for a large number of drive cycles and denote 
it by $\overline{P}(\cos \ta_k)$. Remarkably, we find that 
\bea P_n(\cos \ta_{k_c}) \simeq \overline{P}(\cos \ta_{k_c}) 
\label{twoprob} \eea
for large $n$, where $k_c$ denotes the average momentum of a coarse-grained 
cell that has $N_c$ momenta on the left hand side of Eq.~\eqref{twoprob}, and 
the right hand side is obtained from the motion of a single momentum mode at 
$k_c$. 

For an infinite temperature ensemble, we expect that the $N_c$ points will 
cover the unit sphere uniformly
at large $n$ even though all these points started from the south pole at 
$n=0$ due to the choice of $|\psi(0) \rangle$. Thus, $P_n(\cos \ta_{k_c})$
at large $n$ should be independent of the value of the average momentum of the 
cell, $k_c$, and thus equal 1/2, if the local description is that of an 
infinite temperature ensemble. 
We will derive $\overline{P}(\cos \ta_{k_c})$ for the perfectly periodic case 
in the next subsection and see that numerically $P_n(\cos \ta_{k_c})$ 
approaches $\overline{P}(\cos \ta_{k_c})$ for large $n$. It turns out that 
$\overline{P}(\cos \ta_{k_c})$ depends on $k_c$; this again shows that the 
description is not locally that of an infinite temperature ensemble for 
periodic driving.

\subsection{Probability distribution for periodic driving: $dT=0$}
\label{subsecIIIb}

We will first derive the form of $\overline{P} (\cos \ta_k)$ when there is no 
randomness, 
i.e., when $dT=0$. In this case, the time evolution operator after one time
period $T$ for the momentum mode $k$ (where $0 \le k \le \pi$) is given by
\bea U_k(T) &=& \exp [-i(T/2)\{ 2(g_f - \cos k) \si^z + 2 (\sin k) \si^x \}] 
\non \\
&& \times \exp [-i(T/2)\{ 2(g_i - \cos k) \si^z + 2 (\sin k) \si^x \}]. \non \\
&& \label{uk} \eea
Since $U_k(T)$ is an element of the group $SU(2)$, it can be written in the 
form of Eq.~\eqref{uk2}, where $\ga_k$ and the unit vector $\he_k = (e_{1k}, 
e_{2k},e_{3k})$ can be derived by using the expression given in Eq.~\eqref{uk}.
Then the two-component spinor obtained after $n$ drives is given by
\bea \left( \begin{array}{c}
u_k(n) \\
v_k(n)\end{array} \right) &=& (U_k(T))^n \left( \begin{array}{c}
0 \\
1 \end{array} \right) \non \\
&=& \left( \begin{array}{c}
-i ~\sin (n \ga_k) ~(e_{1k} - i e_{2k}) \\
\cos (n \ga_k) ~+~ i ~\sin (n \ga_k) ~e_{3k} \end{array} \right). \eea
Hence 
\bea \cos \ta_{nk} &=& |u_k(n)|^2 - |v_k(n)|^2 \non \\
&=& - e_{3k}^2 - \cos (2n \ga_k) ~(1- e_{3k}^2), \label{costank} \eea
where we have used the identity $e_{1k}^2 + e_{2k}^2 + e_{3k}^2 = 1$. We now 
assume that $\ga_k/\pi$ is an irrational number; this is justified since 
rational numbers lying in the range $[0,1]$ form a set of measure zero.
Hence the set of numbers $2 n \ga_k ~{\rm mod}~ \pi$ covers the region 
$[0,\pi]$ uniformly as $n \to \infty$. We now use the fact that if 
$2 n \ga_k$ is uniformly distributed from 0 to $\pi$, the probability 
distribution of the variable $v=\cos (2n \ga_k)$ is given by
\beq P(v) ~=~ \frac{1}{\pi ~\sqrt{1 - v^2}} \label{pv} \eeq
for $-1 \le v \le 1$. [This is because $P(v) = (1/\pi) \int_0^\pi ~d\ta ~
\de (v - \cos \ta) = 1/(\pi \sqrt{1-v^2})$.] Using Eq.~\eqref{pv} in 
Eq.~\eqref{costank}, we find that the probability distribution of 
$\cos \ta_k$ is given by
\beq \overline{P}(\cos \ta_k) ~=~ \frac{1}{\pi ~\sqrt{(1 + \cos \ta_k)~ 
(1 - 2 e_{3k}^2 - \cos \ta_k)}} \label{pcostk} \eeq
for $-1 \le \cos \ta_k \le 1 - 2 e_{3k}^2$, while $\overline{P} (\cos \ta_k) =
0$ for $1 - 2 e_{3k}^2 < \cos \ta_k \le 1$. We observe that this distribution
has square root divergences at two points given by $\cos \ta_k = -1$ and
$1 - 2 e_{3k}^2$, and the range of $\cos \ta_k$ is given by $2(1-e_{3k}^2)$. 
Further, the distribution is correctly normalized so that $\int_{-1}^1 
d (\cos \ta_k) \overline{P} (\cos \ta_k) = 1$. 

\begin{figure}
{\includegraphics[width=\hsize]{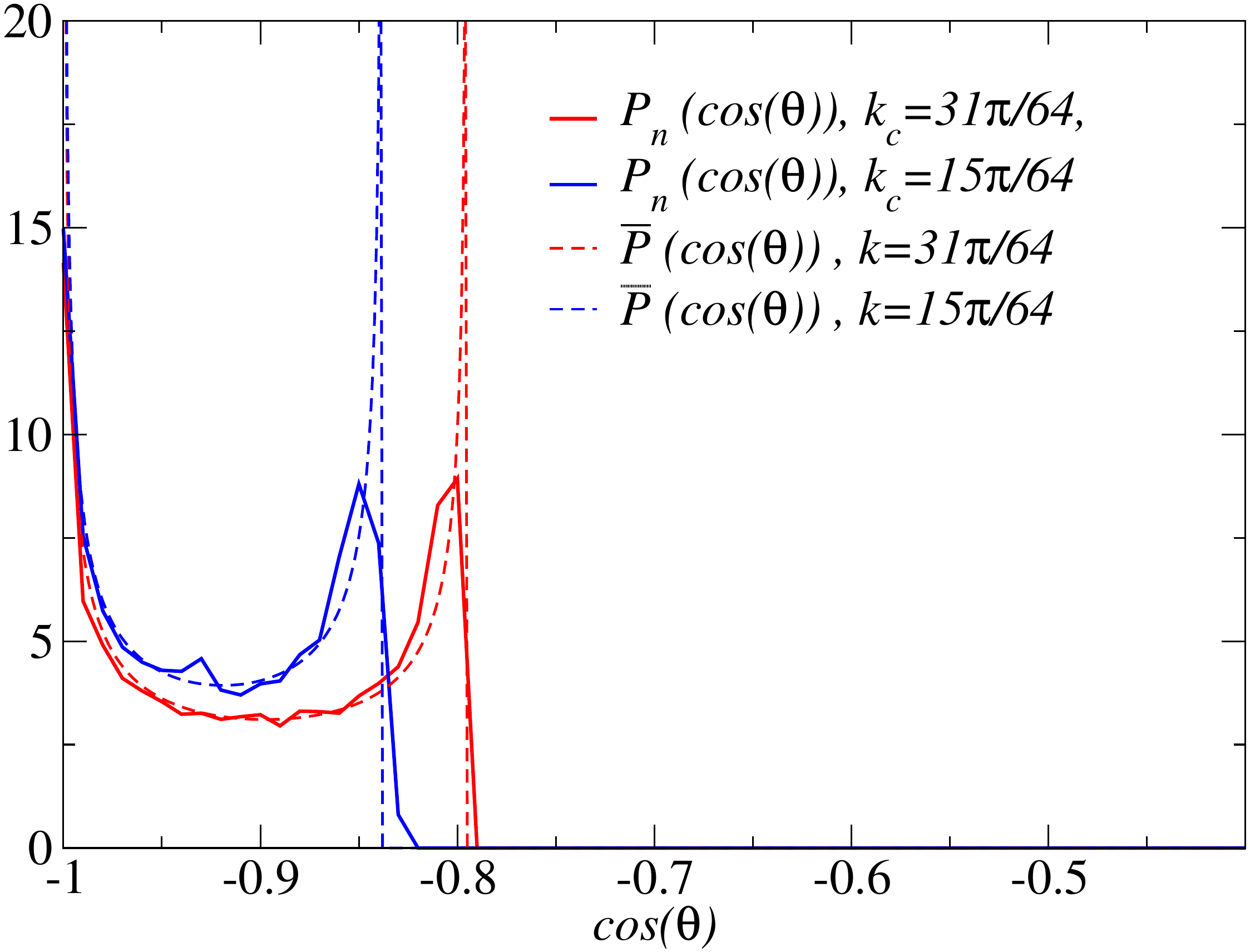}} 
\caption{Plot of $P_n(\cos \ta_k)$ calculated at $n=5 \times 10^5$ 
for a system size of $L=131072$ spins, where the number of coarse-grained 
cells equal $\mathcal{N}_{\mathrm{cell}}=32$ and each cell contains
$N_c=2048$ consecutive momenta. Results for $k_c=31\pi/64$ (red, solid) 
and $k_c=15\pi/64$ (blue, solid) are shown, where $k_c$ denotes the average 
momentum of the corresponding cell. $\overline{P}(\cos \ta_k)$ at 
$k=31\pi/64$ (red, dashed) and $k=15\pi/64$ (blue, dashed) are also shown 
using the analytical result in Eq.~\eqref{pcostk}. The drive parameters
are $g_i=4, g_f=2, T=0.2$ and $dT=0$.} \label{fig2main} \end{figure}

Given the values of the parameters $g_i, ~g_f, ~T$ and $k$, one can calculate
$\he_k$ and therefore $e_{3k}$ from Eqs.~\eqref{uk} and \eqref{uk2}. 
Eq.~\eqref{pcostk} then gives the probability distribution of $\cos \ta_k$ 
which is generated by a large number of drives, as shown in 
Fig.~\ref{fig2main}. We also calculate $P_n(\cos \ta_k)$ for the 1D transverse 
field Ising model with 
$L=131072$ where the positive momenta in the Brillouin zone are divided into 
$\mathcal{N}_{\mathrm{cell}}=32$ cells, each thus containing $N_c=2048$ 
consecutive momenta; we display the result for a large $n$ ($=5 \times 
10^5$) in Fig.~\ref{fig2main} which validates Eq.~\eqref{twoprob}.

Eq.~\ref{twoprob} can be analytically understood as follows. For large 
$n$, we have argued above that the right hand side of Eq.~\ref{twoprob} is 
given by Eq.~\eqref{pcostk} with $k=k_c$. Now we look at the left hand side 
of Eq.~\ref{twoprob}; here we have to consider a large but fixed value of
$n$ and average over a range of momenta $\de k$ around $k=k_c$. We begin with
the expressions 
\bea U_k(T) &=& \exp (-i \ga_k \he_k \cdot \vs), \non \\
U_{k + \de k} (T) &=& \exp (-i \ga_{k + \de k} \he_{k + \de k} \cdot \vs).
\label{ukt} \eea
For $\de k \ll 1$, $\ga_k$ will be close to $\ga_{k + \de k}$ and
$\he_k$ will be close to $\he_{k + \de k}$. However,
\bea (U_k(T))^n &=& \exp (-i n\ga_k \he_k \cdot \vs), \non \\
(U_{k + \de k} (T))^n &=& \exp (-i n\ga_{k + \de k} \he_{k + \de k} \cdot \vs),
\eea
and it is clear that if $n$ is very large, $n \ga_k$ and $n \ga_{k + \de k}$
will not be close to each other. In other words, a small value of 
$\ga_{k + \de k} - \ga_k$ gets amplified as $n$ increases but a small value of
$\he_{k + \de k} - \he_k$ does not get amplified. For $\de k \ll 1$, we can 
therefore make the approximation of setting $\he_k = \he_{k + \de k}$, but we 
cannot set $\ga_k = \ga_{k + \de k}$ if we are interested in the $n \to \infty$
limit. Hence, 
\bea | \psi_k (n) \rangle &=& (U_k(T))^n ~\left( \begin{array}{c}
0 \\
1 \end{array} \right), \non \\
| \psi_{k + \de k} (n) \rangle &=& (U_{k + \de k}(T))^n ~
\left( \begin{array}{c}
0 \\
1 \end{array} \right), \label{psikn} \eea
imply, using Eq.~\eqref{costank}, that
\bea \cos \ta_{nk} &=& - e_{3k}^2 - \cos (2n \ga_k) ~(1- e_{3k}^2), \non \\
cos \ta_{n,k+\de k} &=& - e_{3k}^2 - \cos (2n \ga_{k+\de k}) ~(1- e_{3k}^2).
\eea
Now, we can write $\ga_{k+\de k} - \ga_k = g_k \de k$, where $g_k$ is a number 
of order 1. Suppose that $n$ is large enough that $2n g_k \de k \gg \pi$.
This implies that as $k$ goes over all the momenta in a cell lying in the
range $k_c - \de k/2$ to $k_c + \de k/2$, $2 n \ga_k$ mod $\pi$ will cover 
the range $[0,\pi]$ uniformly. Hence in this cell with an average 
momentum $k_c$, the distribution of the variable $v = \cos (2n \ga_k)$ will 
again be given by Eq.~\eqref{pv}. Arguments similar to the ones leading from 
Eq.~\eqref{costank} to Eq.~\eqref{pcostk} will then show that the left hand
side of Eq.~\eqref{twoprob} is also given by Eq.~\eqref{pcostk}.

We will now derive the value to which $(|u_k(n)|^2)_c$ 
converges as $n \rightarrow \infty$ (Fig.~\ref{fig1main}). It is 
straightforward to see that $1/2-(|u_k(n)|^2)_c = -(1/2)(\cos \theta_k(n))_c$,
where the coarse-grained quantity $(\cos \ta_k(n))_c$ is given by
\bea (\cos \ta_k(n))_c &\equiv& \int_{-1}^1 d (\cos \ta_k) P_n (\cos \ta_k) 
\cos (\ta_k). \eea 
As $n \rightarrow \infty$, using Eq.~\ref{twoprob} and Eq.~\ref{pcostk}, we get
\bea (\cos \ta_k(n))_c &=& \int_{-1}^1 d (\cos \ta_k) \overline{P} 
(\cos \ta_k) \cos (\ta_k) \non \\
&=& - ~e_{3k}^2. \eea
This is in agreement with the results shown in Fig.~\ref{fig1main} for 
$1/2 - (|u_k (n)|^2)_c$ which, in the limit $n \to \infty$, is equal to 
$(1/2) e_{3k_c}^2$.

\section{Results for perturbed Floquet system with random noise}
\label{secIV}

We will now study the unitary dynamics when $dT \ll T$ and the $\tau_i$'s in 
Eq.~\eqref{defsequence} are taken to be any typical realization of a random 
sequence of $\pm 1$'s which are chosen with probability 1/2 each. We again 
look at the behavior of the coarse-grained quantities to understand the nature
of the resulting non-equilibrium steady state. We will see that in spite of 
the integrability of the model, the resulting non-equilibrium steady state 
is locally described by an infinite temperature ensemble as $n \to \infty$, 
unlike the case with $dT=0$.

\begin{figure}
{\includegraphics[width=\hsize]{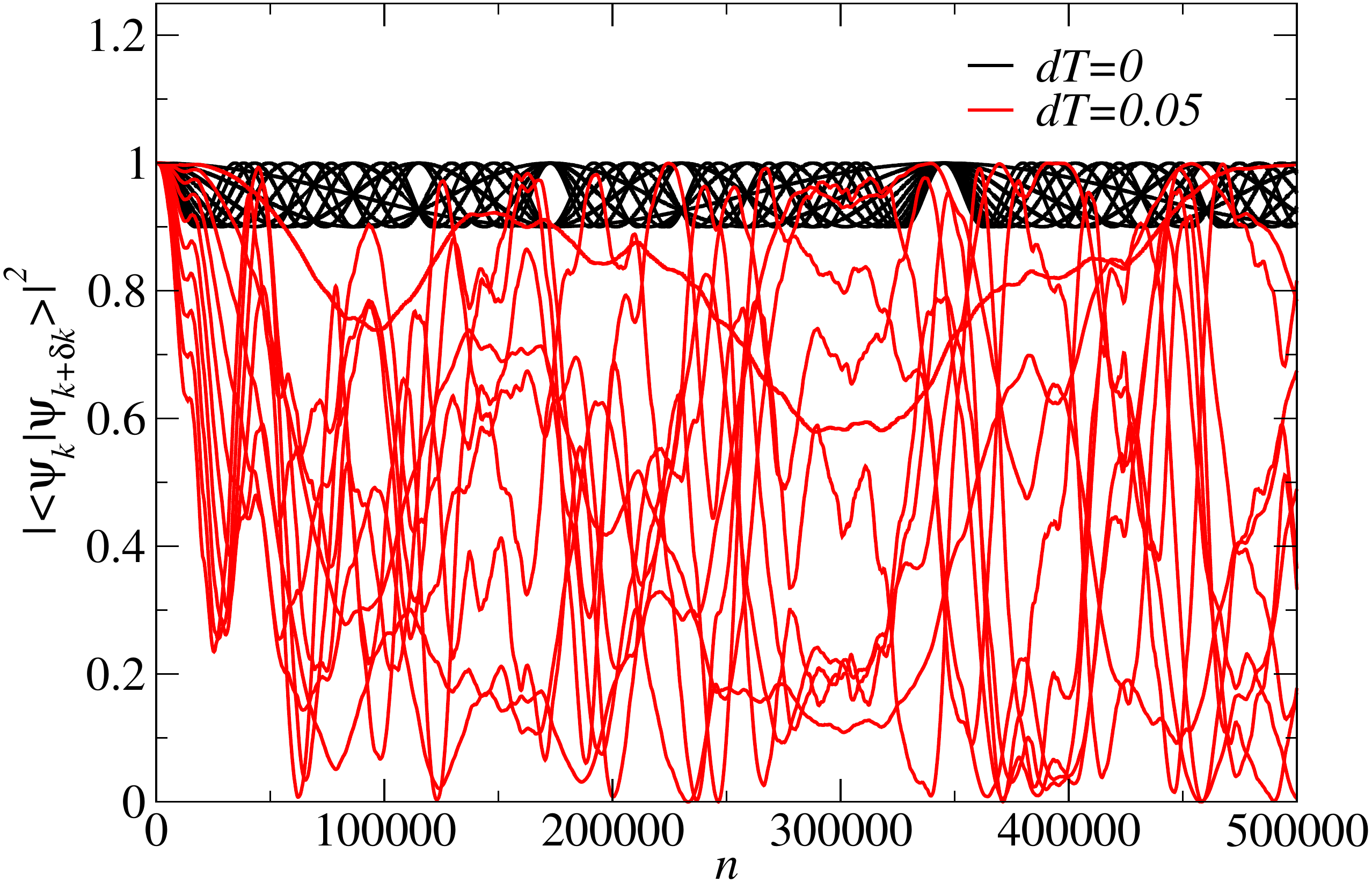}} 
\caption{The overlap $|\langle \psi_k(n)| \psi_{k+\de k}(n) \rangle|^2$ 
for two different kinds of unitary processes as a function of $n$. $dT=0$ 
(black lines) denotes a perfectly periodic drive protocol, while $dT=0.05$ 
(red lines) is a typical realization of a random process where $\tau_i$ are 
chosen to be $\pm 1$ with equal probability (Eq.~\eqref{defsequence}). $T=0.2$
in both cases, and the starting wave function equals $(u_k(0),v_k(0))^T=
(0,1)^T$ for all momenta. $k=\pi/2$ is taken as the reference mode, and its 
overlap calculated with the next ten consecutive momenta are shown where the 
momentum spacing equals $\pi/65536$.} \label{fig3main} \end{figure}

The first quantity we look at is the overlap of a {\it reference} $k$ mode 
$|\psi_k(n) \rangle$ with other $k+\de k$ modes that lie very close to it, 
i.e., where $\de k \ll 1$, as a function of $n$ when the unitary dynamics is 
of the form given in Eq.~\eqref{defsequence} with the $\tau_i$'s taken from 
a typical realization of a random sequence of $\pm 1$'s. 
Fig.~\ref{fig3main} shows a plot
of $|\langle \psi_k (n)|\psi_{k+\de k} (n) \rangle|^2$ versus $n$. At $n=0$, 
all the $|\psi_n(k)\rangle$ start from the same state $(0,1)^T$. We see that
while $|\langle \psi_k (n)|\psi_{k+\de k} (n) \rangle|^2$ stays localized 
between 1 and another value (which is derived in the next paragraph) for the 
perfectly periodic drive with $dT=0$ (see Fig.~\ref{fig3main}), the $dT \neq 0$
case is extremely sensitive to initial conditions; we then find that 
$|\langle \psi_k (n)|\psi_{k+\de k} (n) \rangle|^2$ covers the entire range 
$[0,1]$ at large $n$ when the overlap with different $k+\de k$ modes is 
considered where all $\de k \ll 1$ (see Fig.~\ref{fig3main}). Note that if 
$\de k \ll 1$, then each individual $U_{k+\de k}(T+\tau_n dT)$ is only 
slightly different from $U_k(T+\tau_n dT)$ of the reference $k$ mode. This 
behavior of the overlaps already suggests that the coarse-grained quantities 
behave completely differently in the $dT \neq 0$ case at large $n$.

For the perfectly periodic drive with $dT=0$, we can understand the behavior
of $|\langle \psi_k (n)|\psi_{k+\de k} (n) \rangle|^2$ versus $n$ as follows. 
Following the arguments given in Eqs.~(\ref{ukt}-\ref{psikn}), we find that
\bea && |\langle \psi_k (n)|\psi_{k+\de k} (n) \rangle|^2 \non \\
&& = ~1 ~+~ \sin^2 [n(\ga_k - \ga_{k+ \de k})] ~(e_{3k}^2 ~-~ 1). 
\label{overlap} \eea
For $n \gg 1/|\ga_k - \ga_{k+ \de k}|$, $\sin^2 [n(\ga_k - \ga_{k+ \de k})]$
will oscillate between 0 and 1. Eq.~\eqref{overlap} then implies that 
$|\langle \psi_k (n)|\psi_{k+\de k} (n) \rangle|^2$ will oscillate between
1 and $e_{3k}^2$, and the range of values is given by $1-e_{3k}^2$. We note 
that this is exactly half the range of values of $\cos \ta_k$ which is
given by $2(1-e_{3k}^2)$ according to Eq.~\eqref{pcostk}.

\begin{figure}
{\includegraphics[width=\hsize]{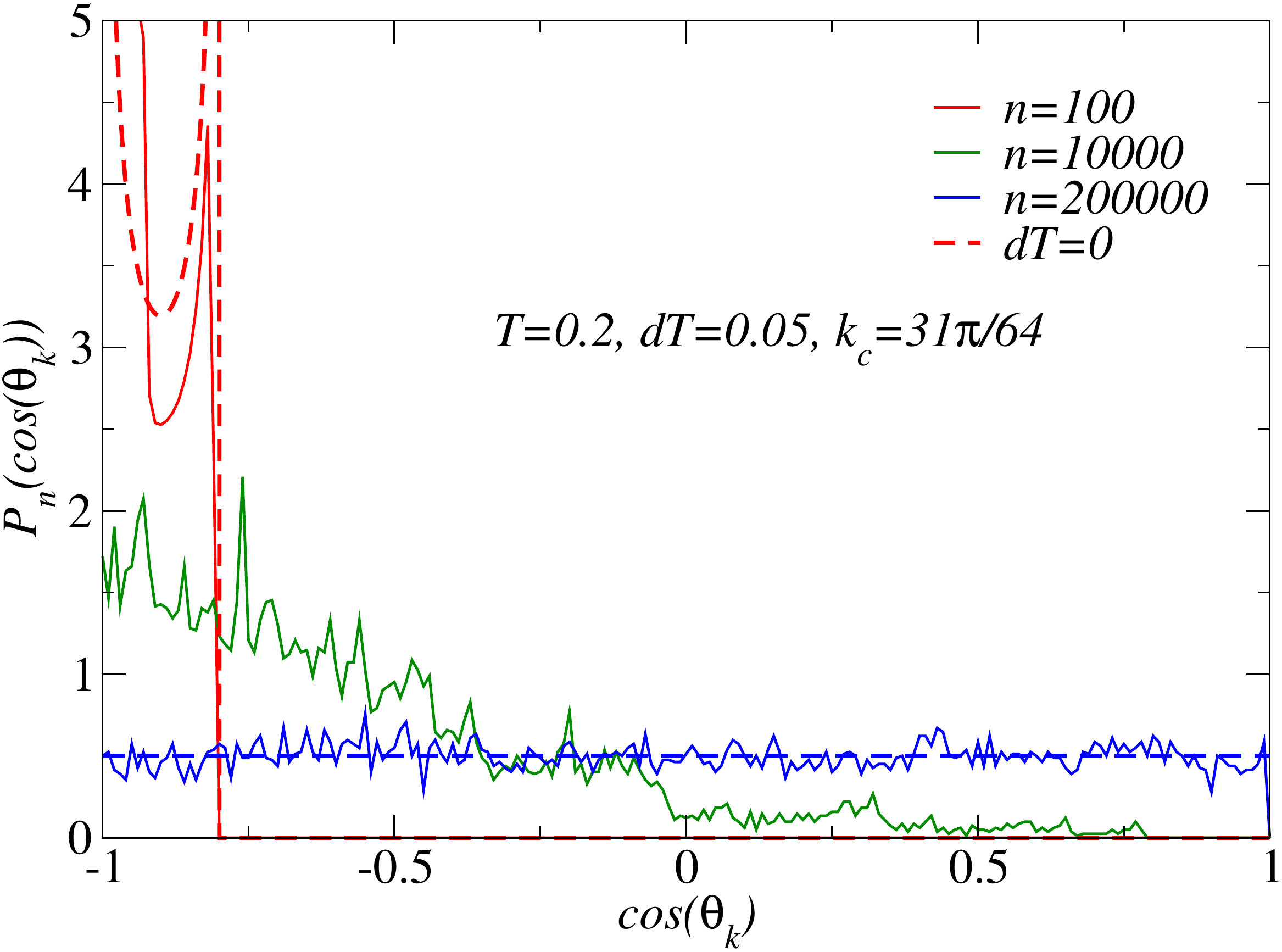}} \\
{\includegraphics[width=\hsize]{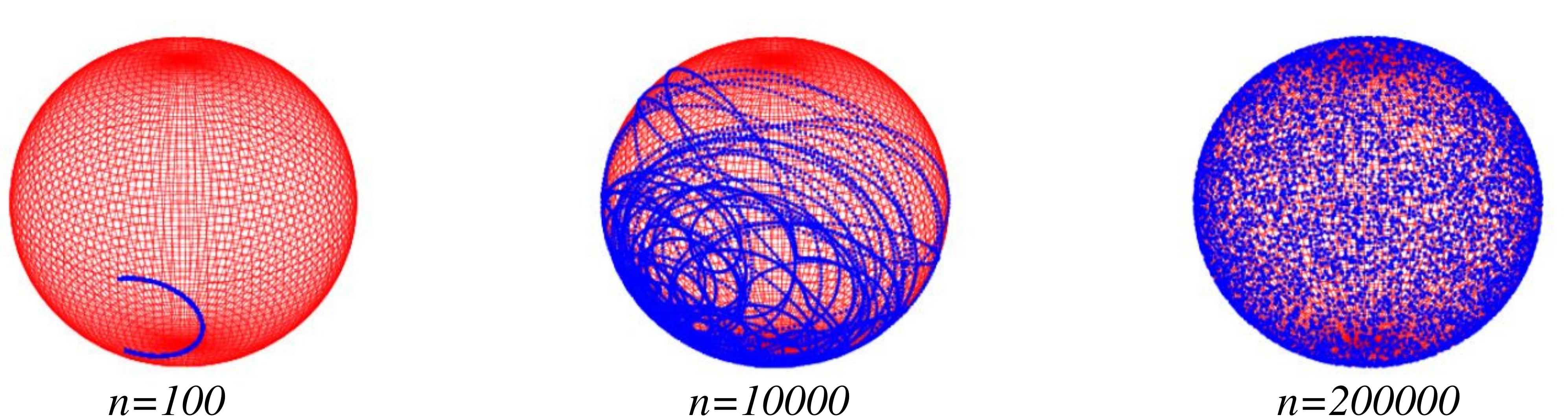}}
\caption{Plots of $P_n(\cos \ta_k)$ for various values of $n$.
The distributions are constructed using $N_c=8192$ points in a 
coarse-grained momentum cell with average momentum $k_c=31\pi/64$; 
the system size equals $L=524288$ and the number of coarse-grained momentum 
cells equal $\mathcal{N}_{\mathrm{cell}}=32$. The drive parameters used are 
$g_i=4, ~g_f=2, ~T=0.2$ and $dT =0.05$. For small $n (=100)$ (red, solid 
line), $P_n(\cos \ta_k)$ approaches the $dT=0$ analytical result in 
Eq.~\eqref{pcostk} (red, dashed line). At larger values of $n (=10000)$ 
(green, solid), the distribution strongly deviates from the $dT=0$ form, and 
it approaches a constant function equal to 1/2 at still larger values of 
$n(=200000)$ (blue, solid). The lower panel shows the positions of the 
$N_c=8192$ points (in blue) on the unit sphere for different values of $n$.} 
\label{fig4main} \end{figure}

We will now study what happens to the probability distribution 
$P_n(\cos \ta_k)$ for the unitary process with $dT \neq 0$. To 
evaluate this quantity, we use a chain with $L=524288$ spins where each 
coarse-grained cell in momentum space contains $N_c=8192$ consecutive momenta 
(hence, $\mathcal{N}_{\mathrm{cell}}=32$) and the spinor at each $k$ equals 
$(0,1)^T$ at $n=0$. The results for one such cell (with average momentum 
$k_c=31\pi/64$) are shown in Fig.~\ref{fig4main}. For small $n$, we see that 
$P_n(\cos \ta_k)$ approaches the distribution for the perfectly periodic 
drive with the same $T$ but with $dT=0$ which has characteristic 
square-root divergences as shown in Eq.~\eqref{pcostk}. However, as $n$ 
increases further, the probability distribution starts deviating strongly 
from this form and approaches a constant distribution eventually 
at large $n$ (see Fig.~\ref{fig4main}); this shows that the $N_c$ 
points that constitute the coarse-grained momentum cell are uniformly spread 
over the unit sphere. The lower panel of Fig.~\ref{fig4main} shows the 
positions of the $N_c$ points on the unit sphere, from which $P_n(\cos 
\theta_k)$ is constructed, for $n=100$, $n=10000$ and $n=200000$. 
We will now provide an understanding of this spreading based
on the idea of random walks.

We define two unitary matrices $U_k(T-dT)$ and $U_k(T+dT)$ given by
\begin{widetext}
\bea U_k(T \pm dT) = \exp [-(i/2)(T \pm dT)\{ (g_f - \cos k) \si^z + (\sin k) 
\si^x \}] \times \exp [-(i/2)(T \pm dT) \{ (g_i - \cos k) \si^z + (\sin k) 
\si^x \}], \non \\
\label{ab0} \eea
\end{widetext}
These are the time evolution operators for the two possible time periods
$T-dT$ and $T+dT$ respectively. The time evolution operator $U_{nk}$ for
$n$ drives is obtained by multiplying $n$ matrices each of which is randomly 
chosen to be either $U_k(T-dT)$ or $U_k(T+dT)$ with probability 1/2 each to 
obtain a particular realization of this random process as shown in 
Eq.~\eqref{defsequence}.

Next, we parameterize the two matrices as
\bea U_k(T-dT) &=& \exp (-i \al_k \ha_k \cdot \vs), \non \\
U_k(T+dT) &=& \exp (-i \be_k \hb_k \cdot \vs). \label{ab0k} \eea
For $dT=0$, these matrices reduce to $U_k(T)$ given in Eq.~\eqref{uk}; 
in that case, we have seen that the evolution operator $U_k(T)$ 
corresponds to a rotation about a unit vector $\he_k$ given in 
Eq.~\eqref{uk2}. If $dT$ is non-zero but small, the unit vector corresponding 
to each drive will have a small random component. To quantify this, we note 
that if $dT$ is small, the unit vectors $\ha_k$ and $\hb_k$ given in 
Eq.~\eqref{ab0k} differ by small angles $\De \phi_{1k} = \arccos 
(\he_k \cdot \ha_k)$ and $\De \phi_{2k} = \arccos (\he_k \cdot \hb_k)$ 
from the unit vector $\he_k$. These two small angles are of the same order 
and we can denote their average as $\De \phi_k$. For $dT=0$ or $\sin k = 0$, 
it is clear from Eqs.~\eqref{uk} and \eqref{ab0} that the unit vectors 
$\he_k, ~\ha_k$ and $\hb_k$ are identical, and therefore $\De \phi_k = 0$. 
Hence, we see that 
\bea \De \phi_k ~\propto~ dT \sin k \label{scalingphik} \eea 
when either $dT$ or $\sin k \to 0$, and Eq.~\eqref{scalingphik} holds 
independently of the exact form of the drive protocol.

If the drives have time 
periods given by $T+dT$ and $T-dT$ randomly, we expect from the theory of 
random walks that after $n$ drives, the unit vector corresponding to $U_{nk}$ 
will differ from $\he_k$ by an angle proportional to $\sqrt{n} \De \phi_k$,
assuming that $\De \phi_k$ is small (this will be true if $dT$ is 
small). However, since the unit vectors lie on a compact space (given by the 
points on the surface of the unit sphere), the deviation of the unit vector of
$U_{nk}$ from $\he_k$ cannot go to $\infty$ as $n \to \infty$; rather, the 
unit vectors will cover the unit sphere uniformly giving rise to a uniform 
probability distribution $\overline{P}(\cos \ta_k) = 1/2$. The behavior of 
the probability distribution $P_n(\cos \ta_k)$ shown in Fig.~\ref{fig4main} 
for large $n$ is thus consistent with Eq.~\eqref{twoprob}.

\begin{figure}
{\includegraphics[width=\hsize]{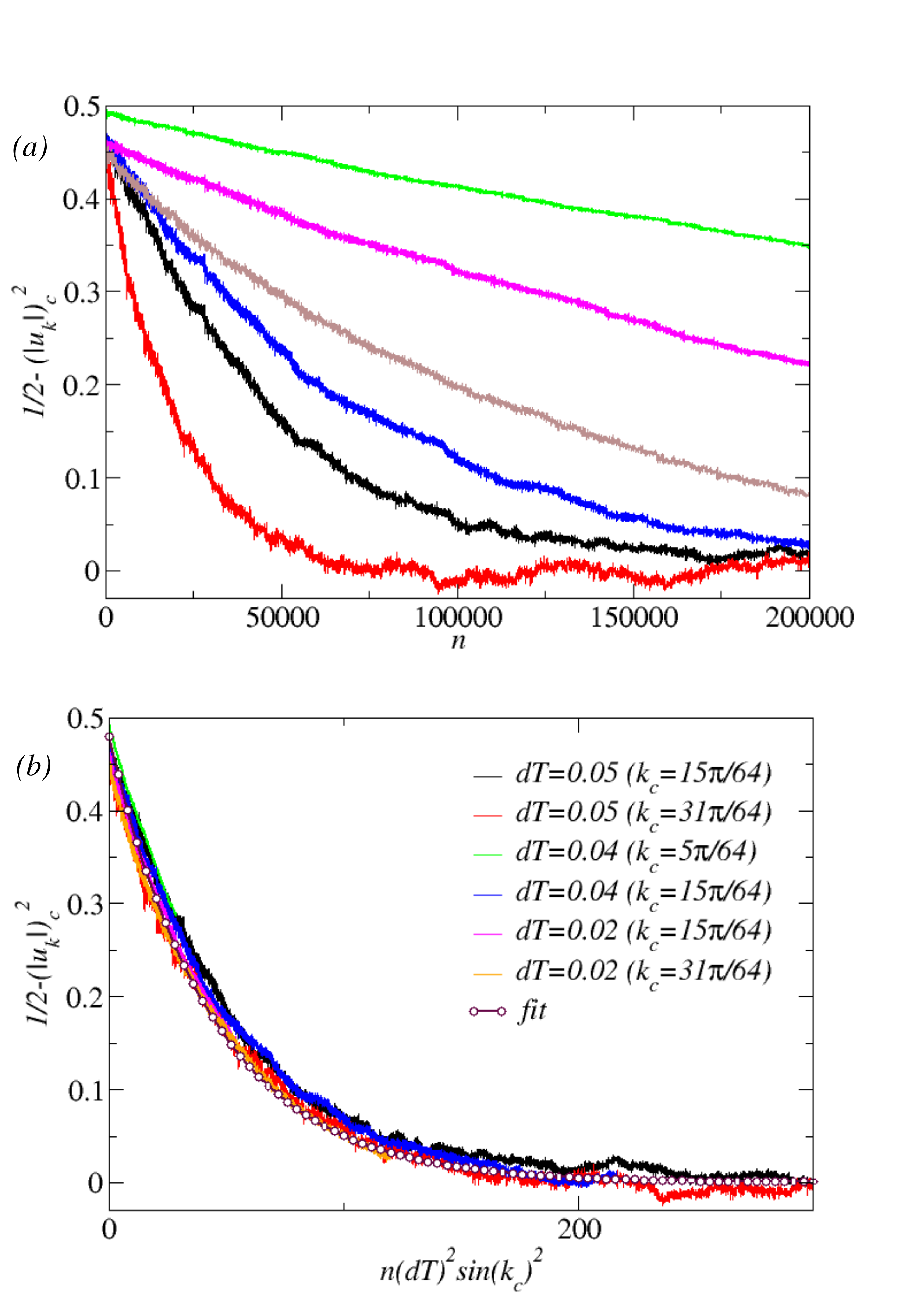}} 
\caption{(a) Behavior of the coarse-grained quantities $1/2-(|u_k(n)|^2)_c$
as a function of $n$ for different values of $dT$ and $k_c$ (average momentum
of the coarse-grained cell). The same random sequence $\tau_i$ has been used
in all the cases. The other parameters are $L=524288$, 
$\mathcal{N}_{\mathrm{cell}}=32$ and $N_c=32$ with $g_i=4,g_f=2$ and $T=0.2$. 
(b) Plot of $1/2-(|u_k(n)|^2)_c$ as a function of $n(dT)^2 (\sin k_c)^2$ 
shows that the decay of $(|u_k(n)|^2)_c$ to 1/2 is consistent with an 
exponential decay with a time scale $\tau_{k_c,dT} = 1/\left[(dT)^2
(\sin k_c)^2\right]$. The fit to an exponential 
function is displayed in (b) using open circles.
The color scheme used in both panels is:
$dT=0.05, k_c=15\pi/64$ (in black), $dT=0.05, k_c=31\pi/64$ (in red), 
$dT=0.04, k_c=5\pi/64$ (in green), $dT=0.04, k_c=15\pi/64$ (in blue), 
$dT=0.02, k_c=15\pi/64$ (in magenta), and $dT=0.02, k_c=31\pi/64$ (in orange).}
\label{fig5main} \end{figure}

Furthermore, from the above argument, the spreading of the overlaps 
$|\langle \psi_k(n)|\psi_{k+\de k} (n) \rangle|^2$ 
as a function of $n$ (as shown in Fig.~\ref{fig3main}) 
when {\it many} neighboring momentum modes very close to the reference mode 
$k$ are considered is controlled by a time scale $nT$, where
$n (\De \phi_k)^2 \simeq n (dT)^2 (\sin k)^2 \sim \mathcal{O}(1)$; hence
the time scale $nT$ varies with the momentum $k$. We explicitly verify 
this by showing the behavior of the corresponding coarse-grained variables 
$(|u_k(n)|^2)_c$ (defined in Eq.~\eqref{coarsegrain1})
in Fig.~\ref{fig5main}. The data is consistent with an exponential decay to 
1/2; however, the rate of decay is clearly sensitive to the values of both 
$dT$ and $k_c$ (the average momentum of the coarse-grained cell) as we 
see in Fig.~\ref{fig5main} (a). However, when the data is plotted versus 
$n(dT)^2(\sin k_c)^2$ (see Fig.~\ref{fig5main} (b)) instead of $n$, we clearly
see that the time scale of the exponential decay to 1/2 is controlled by 
$\tau_{k_c,dT} = 1/\left[(dT)^2(\sin k_c)^2\right]$ as predicted by the 
random walk argument. We have further checked that the coarse-grained 
quantity $(u^*_k(n)v_k(n))_c$ (Eq.~\eqref{coarsegrain1}) relaxes to zero 
for large $n$ for any value of $k_c$. 

The behaviors of both $P_n(\cos \ta_k)$ in Fig.~\ref{fig4main} and 
$(|u_k(n)|^2)_c$ in Fig.~\ref{fig5main} as functions of $n$ show the 
{\it irreversible} approach to an infinite temperature ensemble when local 
quantities are probed.
Finally, we consider the implication of the dependence of the time scale 
$\tau_{k,dT}=1/\left((dT)^2(\sin k_c)^2\right)$ (that controls the relaxation 
of $(|u_k(n)|^2)_c$) on $dT$ and $k_c$ for the rate at which the trace distance
of the reduced density matrix of a subsystem of $l$ consecutive spins from an 
infinite temperature ensemble (using the definition $\mathcal{D}_n(l)$ in 
Eq.~\eqref{distancemeasure}) 
approaches zero after a large number of drives $n$. Since the trace distance 
receives a contribution from all momenta $k$, it should approach zero as 
\beq \mathcal{D}_{n,ITE}(l) ~\sim~ \int_0^\pi dk ~\exp (-n/\tau_{k,dT}), 
\label{trace} \eeq
where $\mathcal{C}_{\infty}(l)$ in Eq.~\eqref{distancemeasure} is calculated 
using an infinite temperature ensemble. 

\begin{figure}
{\includegraphics[width=\hsize]{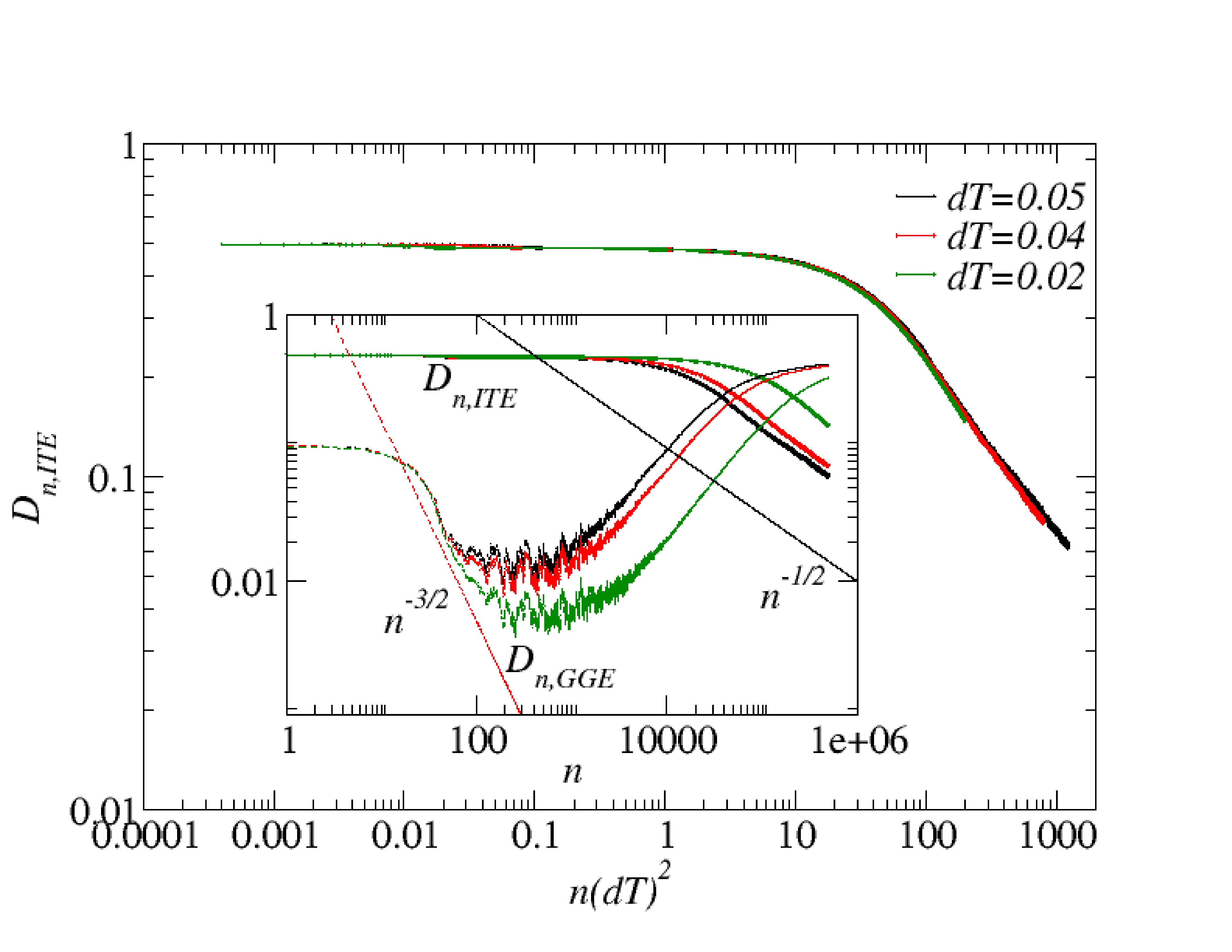}} 
\caption{Plot of $\mathcal{D}_{n,ITE}$ as a function of $n(dT)^2$ for 
different values of $dT$ ($dT=0.05$ in black, $dT=0.04$ in red, and $dT=0.02$ 
in green), where the other drive parameters are $g_i=4, g_f=2$
and $T=0.2$. The system size equals $L=524288$ spins and the subsystem 
consists of $l=16$ consecutive spins. The same random sequence $\tau_i$ has 
been used in all the cases. (Inset) Plot of $\mathcal{D}_{n,GGE}$ as a 
function of $n$.} \label{fig6main} \end{figure}

For large $n$, the integral in \eqref{trace} is dominated by the regions in 
$k$ where $\tau_{k,dT}$ is large, namely, $k=0$ and $\pi$ where $\sin k \to 
0$. Absorbing $n$ in $k$ (or $\pi -k$) in the exponential, we see that 
the trace distance will scale with $n$ as $1/(\sqrt{n} dT)$ when $dT$ is 
small. Thus, $\mathcal{D}_{n,ITE}(l) \sim \mathcal{F}_l(n(dT)^2)$, where 
$\mathcal{F}_l(x) \sim 1/\sqrt{x}$ for $x \gg 1$. Secondly, $\mathcal{F}_l(x)
\sim \mathcal{O}(1)$ when $x \ll 1$ since in this limit the system must relax 
to the p-GGE that emerges for $dT=0$. Thus, the prethermal regime 
where the system initially approaches the p-GGE exists for time scales that 
depend on $dT$ as $n \sim 1/(dT)^2$ for $dT \ll T$.
The numerical data for $dT \neq 0$ for the behavior of $\mathcal{D}_{n,ITE}
(l)$ as a function of $x =n(dT)^2$ is consistent with both these expectations, 
i.e., $\mathcal{F}_l \sim 1/\sqrt{x}$ for $x \gg 1$ and $\mathcal{F}_l \sim 
\mathcal{O}(1)$ for $x \ll 1$, as shown in Fig.~\ref{fig6main} for a fixed 
$T=0.2$ and different values of $dT$. It is also instructive to look at 
$\mathcal{D}_{n,GGE}$ for non-zero values of $dT$ (Fig.~\ref{fig6main}, 
inset) where $\mathcal{C}_\infty(l)$ is calculated 
using the $dT=0$ p-GGE as a function of $n$. For time scales 
$n \sim 1/(dT)^2$, $\mathcal{D}_{n,GGE}$ decreases as a function of $n$ which 
shows that the system approaches the p-GGE in the prethermal regime. However, 
at larger values of $n$, $\mathcal{D}_{n,GGE}$ 
starts increasing as a function of $n$ which explicitly shows that the 
large-$n$ steady state is not described by a p-GGE. The initial relaxation to a 
p-GGE in the prethermal regime scales either as $\mathcal{D}_{n,GGE} \sim 
n^{-3/2}$ (the case shown in Fig.~\ref{fig6main}, inset) or as 
$\mathcal{D}_{n,GGE} \sim n^{-1/2}$ depending on which dynamical phase the 
periodic drive protocol ($dT=0$) belongs to~\cite{SenNS2016}. 

For completeness, we point out that Ref.~\onlinecite{MarinoS2012} (see also 
Ref.~\onlinecite{RooszJI2016}) also considered a noisy 1D transverse field 
Ising model though not periodically driven on average, and showed that the 
asymptotic steady state is an infinite temperature ensemble. However, they 
considered the behavior of quantities averaged over different realizations of 
noise which resembles an open quantum system with non-unitary dynamics. Here, 
we show that if one restricts to understanding only {\it local} quantities 
as functions of $n$ in a perturbed Floquet integrable system, even the unitary
dynamics of a single typical realization of a random sequence leads to an 
infinite temperature ensemble.

\section{Results for perturbed Floquet system with scale-invariant noise}
\label{secV}

\begin{figure}
{\includegraphics[width=\hsize]{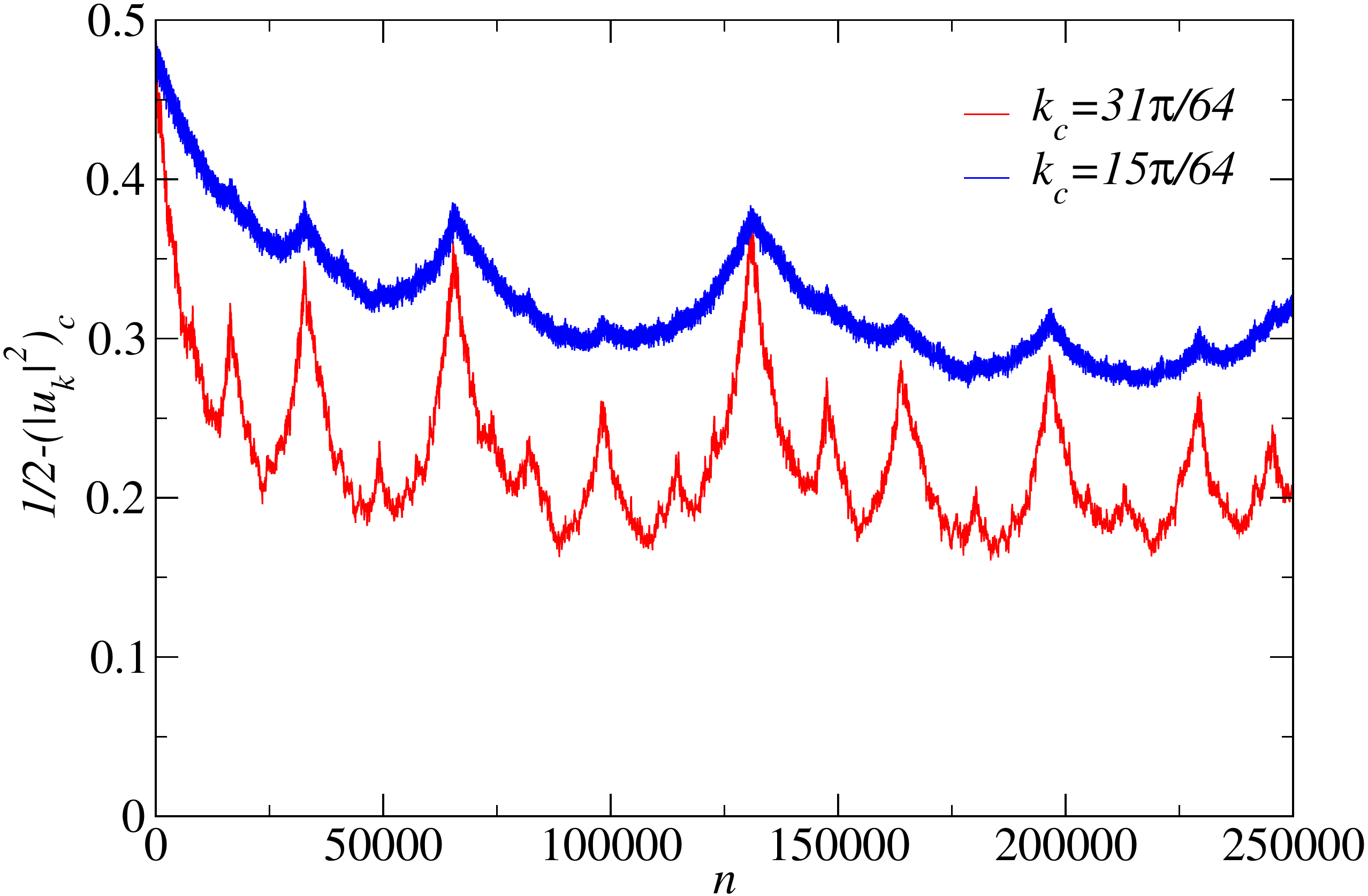}} 
\caption{Behavior of the coarse-grained quantities $(|u_k(n)|^2)_c$ as a 
function of $n$ when the $\tau_i$'s are chosen according to the Thue-Morse 
sequence. 
The parameters used are $L=131072$, $g_i=4$, $g_f=2$, $T=0.2$ and $dT=0.05$. 
Each coarse-grained cell contains $N_c = 2048$ consecutive momenta and $k_c$ 
denotes the average momentum of such a cell. Results for $k_c=31\pi/64$ (lower
red curve) and $k_c=15\pi/64$ (upper blue curve) are shown.} 
\label{fig7main} \end{figure}

We will now study the unitary dynamics when $dT \neq 0$ but the $\tau_i$'s 
in Eq.~\eqref{defsequence} are not a random sequence of $\pm 1$ but are instead 
given by the scale-invariant Thue-Morse sequence in Eq.~\eqref{TMS} that is 
neither periodic nor random. The average time-period of such a drive protocol 
continues to equal $T$. To this end, we first show the coarse-grained quantity 
$(|u_k(n)|^2)_c$ in Eq.~\eqref{coarsegrain1} as a function of $n$ where 
the $\tau_i$'s are chosen to be $\pm 1$ according to the Thue-Morse, 
with $T=0.2$ and $dT=0.05$ (Fig.~\ref{fig7main}). We see 
that these quantities behave entirely differently from either the periodic 
case ($dT=0$) in Fig.~\ref{fig1main} or the randomly perturbed case 
in Fig.~\ref{fig5main}. From Fig.~\ref{fig7main}, it is not even clear whether 
a well-defined non-equilibrium steady state exists as $n \to \infty$. We will 
show below that in fact a non-equilibrium steady state does exist in the 
large-$n$ limit if the local quantities are 
observed not as a function of $n$, but instead as $2^n$ (hence, geometrically 
instead of stroboscopically). New {\it emergent} conserved quantities, similar
to Eq.~\eqref{conservedIk} when $2^n$ is used instead of $n$, appear in this 
unitary dynamics but only when $n$ becomes sufficiently large, which then allow
for the construction of a ``geometric'' GGE for the resulting steady state.

We begin by noting that the evolution operators for drives corresponding 
to $T-dT$ and $T+dT$ at a particular $k$ are respectively given by the 
matrices $U_k(T-dT)$ and $U_k(T+dT)$ given in
Eqs.~\eqref{ab0}; for simplicity of notation, we will not write the
subscript $k$ in this section henceforth. Furthermore, we will denote 
$U(T-dT)$ by $A_0$ and $U(T+dT)$ by $B_0$ here. Then the operators 
corresponding to the pair of drives $(\tau_i,\tau_{i+1})$ = $-1,+1$ and 
$(\tau_i,\tau_{i+1})$=$+1,-1$ are given by $A_1 = B_0 A_0$ and $B_1 =
A_0 B_0$ respectively. We now focus on the case where $n=2^m$ where 
$m=0,1,2,\cdots$. It is then easy to see that the unitary dynamics can be 
entirely expressed in terms of the new matrices $A_1$ and $B_1$. We 
illustrate this in the second line of Eq.~\eqref{coarsegrain_in_time} for 
the case of $m=4$ (thus $n=16$). 
\begin{widetext}
\bea \boxed{A_0 B_0} \boxed{B_0 A_0} \boxed{B_0 A_0} \boxed{A_0 B_0} 
\boxed{B_0 A_0} \boxed{A_0 B_0} \boxed{A_0 B_0} \boxed{B_0 A_0} &|&\psi_k(n=0)
\rangle \non \\
 \boxed{B_1 A_1} \boxed{A_1 B_1} \boxed{A_1 B_1} \boxed{B_1 A_1} &|&
\psi_k(n=0) \rangle \non \\
 \boxed{A_2 B_2} \boxed{B_2 A_2} &|&\psi_k(n=0) \rangle \non \\
 \boxed{B_3 A_3} &|&\psi_k(n=0) \rangle \non \\
 A_4 &|&\psi_k(n=0) \rangle \label{coarsegrain_in_time} \eea 
\end{widetext} 
This motivates us to define matrices $A_m$ and $B_m$ recursively as
\bea A_{m+1} &=& B_m ~A_m, \non \\
B_{m+1} &=& A_m ~B_m, \label{abn1} \eea
for all $m \ge 0$. We can then show that the evolution operator after exactly 
$2^m$ drives is given by $A_m$. (Eq.~\eqref{coarsegrain_in_time} illustrates 
this for $m=4$, i.e., $n=16$). Note that this provides a very efficient method
to calculate the dynamics after $n=2^m$ which only requires $\mathcal{O}(m)$ 
matrix multiplications at each $k$ instead of $\mathcal{O}(2^m)$. {\it 
We will show that remarkably $A_m=B_m$ when $m$ becomes sufficiently large
and thus there is an emergent periodicity when the system 
is viewed geometrically (i.e., after every $n=2^m$).}

We will now study how the matrices $A_m, ~B_m$ defined in Eqs.~\eqref{abn1}
behave in the limit $m \to \infty$. To this end, we parameterize these 
matrices as
\bea A_m &=& \exp (-i \al_m \ha_m \cdot \vs), \non \\
B_m &=& \exp (-i \be_m \hb_m \cdot \vs). \label{abn2} \eea
We also define the angle $\phi_m$ between the unit vectors $\ha_m$ and
$\hb_m$, namely, 
\beq \phi_m ~=~ \arccos (\ha_m \cdot \hb_m), \label{phn1} \eeq
with $0 \le \phi_m \le \pi$.
Using Eqs.~(\ref{abn1}-\ref{phn1}), we find that
\bea \cos \al_{m+1} &=& \cos \be_{m+1} \non \\
&=& \cos \al_m \cos \be_m ~-~ \sin \al_m \sin \be_m \cos \phi_m, \non \\
\sin \al_{m+1} ~\ha_{m+1} &=& \sin \al_m \cos \be_m ~\ha_m ~+~ \sin \be_m 
\cos \al_m ~\hb_m \non \\
&& +~ \sin \al_m \sin \be_m ~\ha_m \times \hb_m, \non \\
\sin \be_{m+1} ~\hb_{m+1} &=& \sin \al_m \cos \be_m ~\ha_m ~+~ \sin \be_m 
\cos \al_m ~\hb_m \non \\
&& -~ \sin \al_m \sin \be_m ~\ha_m \times \hb_m. \label{recur} \eea
{}From Eqs.~\eqref{recur}, we can prove the following results. First, we
find that $\al_m = \be_m$ for all $m \ge 1$. We will therefore use only the
variable $\al_m$ henceforth. Next, we find the recursion relations
\bea \cos \al_{m+1} &=& \cos^2 \al_m ~-~ \sin^2 \al_m ~\cos \phi_m, 
\label{aln} \\
\tan \left(\frac{\phi_{m+1}}{2}\right) &=& |\tan \al_m| ~\sin \left(
\frac{\phi_m}{2}\right). \label{phn2} \eea

Since $\al_m = \be_m$ for all $m \ge 1$, Eqs.~\eqref{recur} imply that
\beq \ha_{m+1} ~+~ \hb_{m+1} ~=~ \frac{\sin (2 \al_m)}{\sin \al_{m+1}} ~
(\ha_m ~+~ \hb_m). \label{ambm} \eeq
This means that the unit vector which points in the direction of $\ha_m ~+~
\hb_m$ is the same for $m$ and $m+1$, up to a sign; hence, up to a sign, the 
direction of this unit vector does not change at all with $m$, all the way 
from $m=1$ to $\infty$. These arguments will break down if either $\ha_m$
or $\hb_m$ is ill-defined; according to Eq.~\eqref{abn2}, this can only happen
if $\al_m$ or $\be_m$ becomes exactly equal to 0 or $\pi$. However these cases 
require very special choices of $A_0$ and $B_0$ which form a set of measure 
zero; we will therefore ignore such special cases. 

To summarize, we find that the dynamics given by Eq.~\eqref{abn1} 
conserves two quantities at every iteration: $\alpha_m = \beta_m$, and the 
direction of the unit vector ${\hat a}_m + {\hat b}_m$ is conserved for all 
$m \ge 1$. The existence of these conserved quantities considerably 
simplifies the analysis as we will see below.

\begin{figure}
{\includegraphics[width=\hsize]{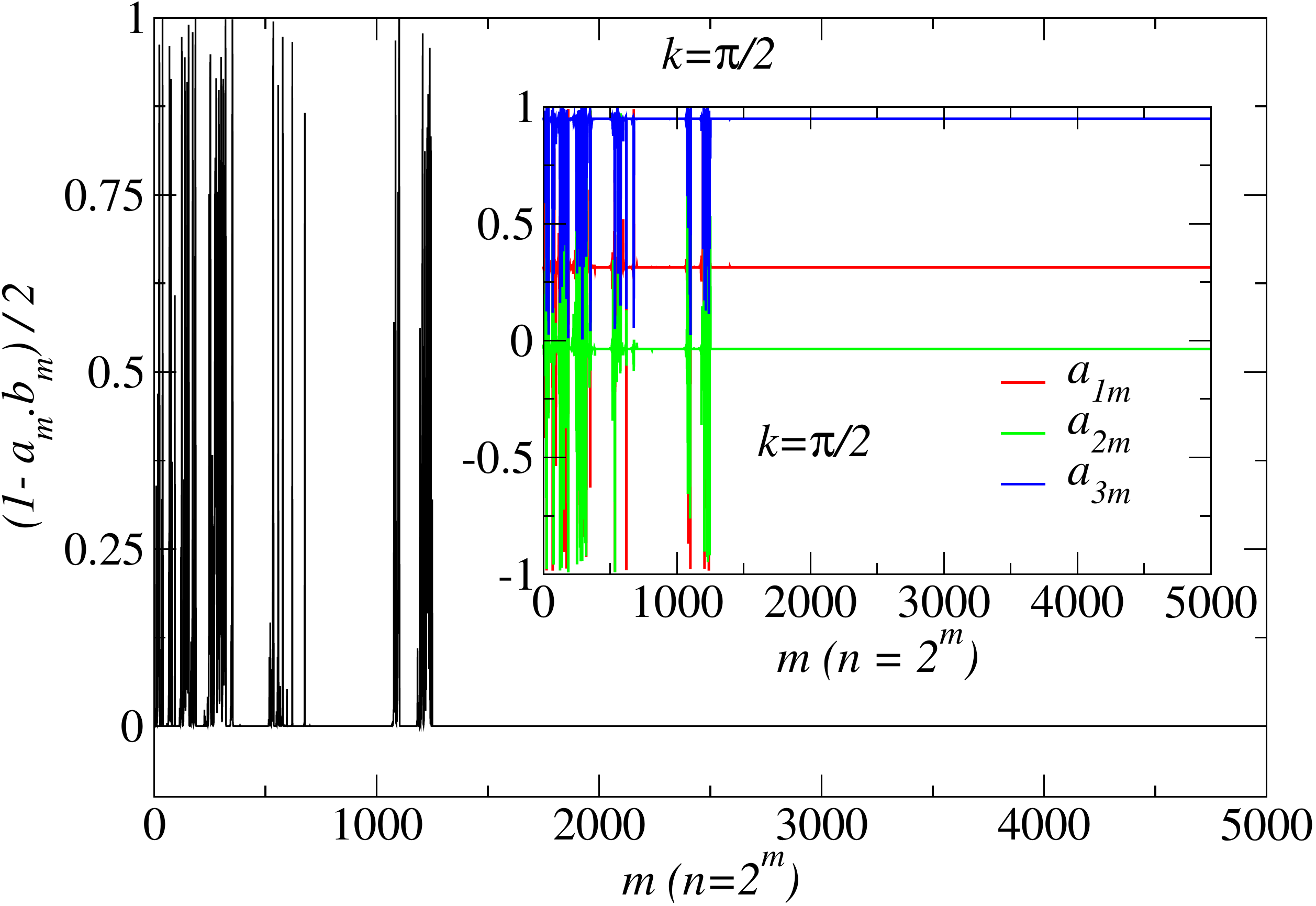}} 
\caption{Behavior of $\hat{a}_m \cdot \hat{b}_m$ as a function of $m$ for a 
unitary process with $k=\pi/2$, $T=0.2$ and $dT=0.05$, where the $\tau_i$'s are
chosen according to the Thue-Morse 
sequence. (Inset) Behaviors of the three components 
of $\hat{a}_m$ ($a_{1m}$ (red), $a_{2m}$ (green), $a_{3m}$ (blue)) as functions
of $m$ for the same unitary process. The unit vector $\hat{a}_m$ points in the 
direction of $\hat{a}_1+\hat{b}_1$ for large $m$ (from Eq.~\ref{ambm}) since 
$\hat{a}_m=\hat{b}_m$ for $m \gg 1$.} \label{fig8main} \end{figure}

For sufficiently large $m$, we find numerically that $\phi_m \to 0$ 
although $\phi_m$ does not approach zero 
monotonically (see Fig.~\ref{fig8main}). We can understand
this as follows. If $\phi_m \to 0$, Eq.~\eqref{aln} implies that $\al_{m+1}
\to 2 \al_m ~{\rm mod}~ \pi$. Assuming that $\al_m/\pi$ is irrational, we see 
that $\al_m$ will cover the region $[0,\pi]$ uniformly as $m \to \infty$.
Next, assuming $\phi_m, ~\phi_{m+1}$ to be small, we obtain 
\beq \phi_{m+1} ~=~ |\tan \al_m| ~\phi_m \label{phn3} \eeq
from the lowest order expansion of the terms appearing in Eq.~\eqref{phn2}.
Since $\al_m$ covers $[0,\pi]$ uniformly, $|\tan (\al_m)|$ varies all the
way from zero to $\infty$; this explains why $\phi_m$ does not
monotonically approach zero. In fact, 
\beq \frac{\phi_{m+N}}{\phi_m} ~=~ \exp [\sum_{j=0}^{N-1} 
\log (|\tan \al_{m+j}|)], \eeq
and for large $N$, a uniform distribution of $\al_m$ implies that 
$\sum_{j=0}^{N-1} \log (|\tan \al_{m+j}|) \to (N/\pi) \int_0^\pi d\al \log 
(|\tan \al|)$ which is equal to zero. Thus we would expect that as
$m \to \infty$, $\phi_m$ should stay at a constant value if we stopped at 
the first order expression in Eq.~\eqref{phn3}. We therefore have 
to go to the next order in the expansion of Eq.~\eqref{phn2}. We find that
\beq \phi_{m+1} ~=~ |\tan \al_m| ~\phi_m ~\left(1 ~-~ \frac{\phi_m^2}{24} 
\left( 1~+~2 \tan^2 \al_m \right) \right). \label{phn4} \eeq
The fact that the last factor on the right hand side is always less
than 1 explains why $\phi_m$ eventually goes to zero as $m \to \infty$;
this explains the numerical behavior shown in Fig.~\ref{fig8main}.

As $m \to \infty$, the fact that $\phi_m \to 0$ means that
$\ha_m - \hb_m \to 0$. The discussion following Eq.~\eqref{ambm} then implies
that $\ha_\infty = \hb_\infty$ is equal, up to a sign, to the unit vector
which points in the direction of $\ha_1 + \hb_1$. Hence the value of 
$\ha_\infty = \hb_\infty$ can be found right in the beginning when we 
calculate $\ha_1$ and $\hb_1$.

We now examine what happens if we view the system geometrically after $n=2^m$
drives where $m=0,1,2,\cdots$. In Fig.~\ref{fig8main} (inset), we show the 
components of the unit vector $\ha_m$ versus $m$ for $g_i=4, ~g_f=2, ~T=0.2, ~
dT =0.05$, and $k=\pi/2$. (In that figure, we have taken the third component
of $\ha_m$, called $a_{3m}$, to be non-negative for all values of $m$. 
According to Eq.~\eqref{abn2}, this can always be ensured,
by flipping $\al_m \to - \al_m$ and $\ha_m \to - \ha_m$ if necessary.)
We see that for $m$ up to about 1300, there are intermittent large 
fluctuations in both $\ha_m$ and $\hb_m$. As Eq.~\eqref{phn3} shows, these 
fluctuations occur when $|\tan \al_m|$ becomes large, namely, when $\al_m$ 
comes close to $\pi/2$. For very large values of $m$, $\ha_m$ and $\hb_m$ 
settle down to the same value (Fig.~\ref{fig8main}, main panel). We therefore 
conclude that after an extraordinarily large number of drives given by $2^m$, 
where $m \gtrsim 1300$ (we have checked this for other values of $k$ also), 
$\ha_m$ and $\hb_m$ become identical and independent of $m$. 

Since the unitary 
dynamics for $n=2^m$ can be represented by a single unitary matrix $A_m$ for 
the Thue-Morse sequence, this immediately shows that $\mathcal{J}_k$ defined as
\bea \mathcal{J}_k (2^m)&=& \langle \psi_k(2^m)|\hat{a}_\infty \cdot \vec{\si}|
\psi_k(2^m) \rangle \non \\
&=& \langle \psi_k(0)|A_m^\da (\hat{a}_\infty \cdot \vec{\si}) A_m |
\psi_k(0) \rangle \non \\
&=& \langle \psi_k(0)|e^{i\al_m \hat{a}_m \cdot \vec{\si}} (\hat{a}_\infty 
\cdot \vec{\si}) e^{-i\al_m \hat{a}_m \cdot \vec{\si}} |\psi_k(0) 
\rangle \non \\
\label{conservedJk} \eea 
equals $\mathcal{J}_k \equiv \langle \psi_k(0)|\hat{a}_\infty \cdot \vec{\si} |
\psi_k(0) \rangle$ 
and becomes independent of $m$ for sufficiently large $m$ at each momentum 
$k$. We should stress here, that unlike $\mathcal{I}_k$ defined in 
Eq.~\eqref{conservedIk} for the $dT=0$ case which is conserved as a function of 
$n$, the quantities $\mathcal{J}_k$ are conserved geometrically, i.e., when 
$n=2^m$. Furthermore, one requires $m$ to be larger than a threshold value 
($m \gtrsim 1300$ for the drive parameters shown in Fig.~\ref{fig8main}) for 
these conserved quantities to emerge at all momenta $k$, and these conserved 
quantities are not manifest in the unitary dynamics below this 
threshold value of $m$. We have also verified numerically that $\hat{a}_m$ 
and $\hat{b}_m$ become identical and independent of $m$ at large $m$ not 
only for $dT \ll T$ but for any value of $dT/T$.

\begin{figure}
{\includegraphics[width=\hsize]{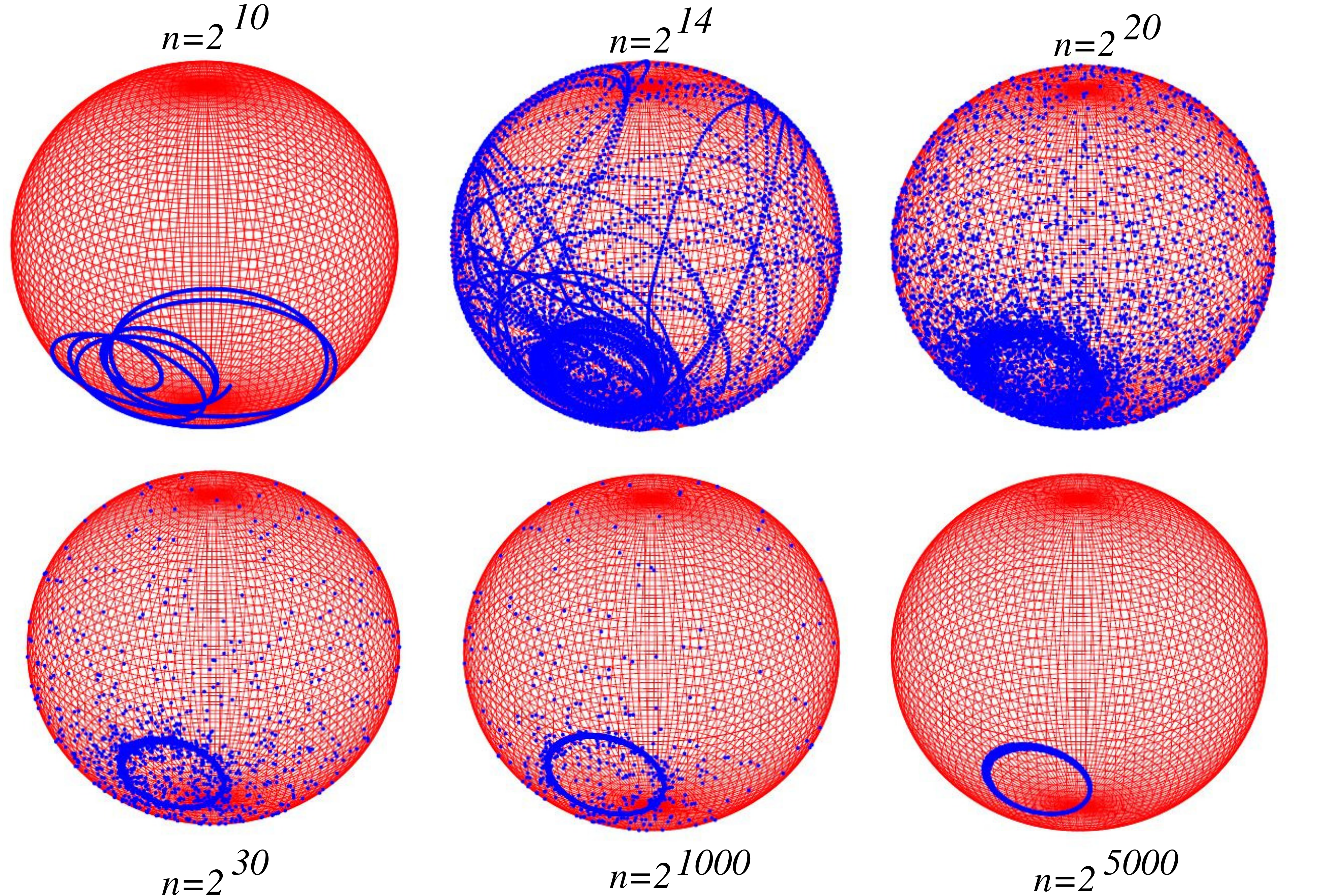}} 
\caption{The positions of the $N_c$ points (in blue) contained in a 
coarse-grained momentum cell on the unit sphere for various values of $n=2^m$, 
where the system is viewed geometrically, and the $\tau_i$'s are chosen 
according to the Thue-Morse sequence. The parameters used are $L=131072$, $g_i
=4$, $g_f=2$, $T=0.2$ and $dT=0.05$. Each coarse-grained momentum cell contains
$N_c=2048$ consecutive momenta and the results displayed are for a cell with 
average momentum of $k_c=31\pi/64$.} \label{figpanelTM} \end{figure}

The motion of the $N_c (\gg 1)$ consecutive momenta contained 
in any of the coarse-grained momentum cells on the unit sphere as a function 
of $n$ provides another way to geometrically see the emergent conserved 
quantities of this unitary dynamics (with the $\tau_i$'s chosen according to 
the Thue-Morse sequence). As shown in Fig.~\ref{figpanelTM}, when the system 
is viewed geometrically, i.e., with $n=2^m$, the $N_c$ points that belong to a
momentum cell with average momentum $k_c$ initially perform a complicated 
dynamics on the unit sphere but eventually settle down to a very simple motion 
along a circle in the plane defined by $\hat{a}_\infty$ (at $k_c$) for large 
enough values of $m$ beyond a particular threshold.

\begin{figure}
{\includegraphics[width=\hsize]{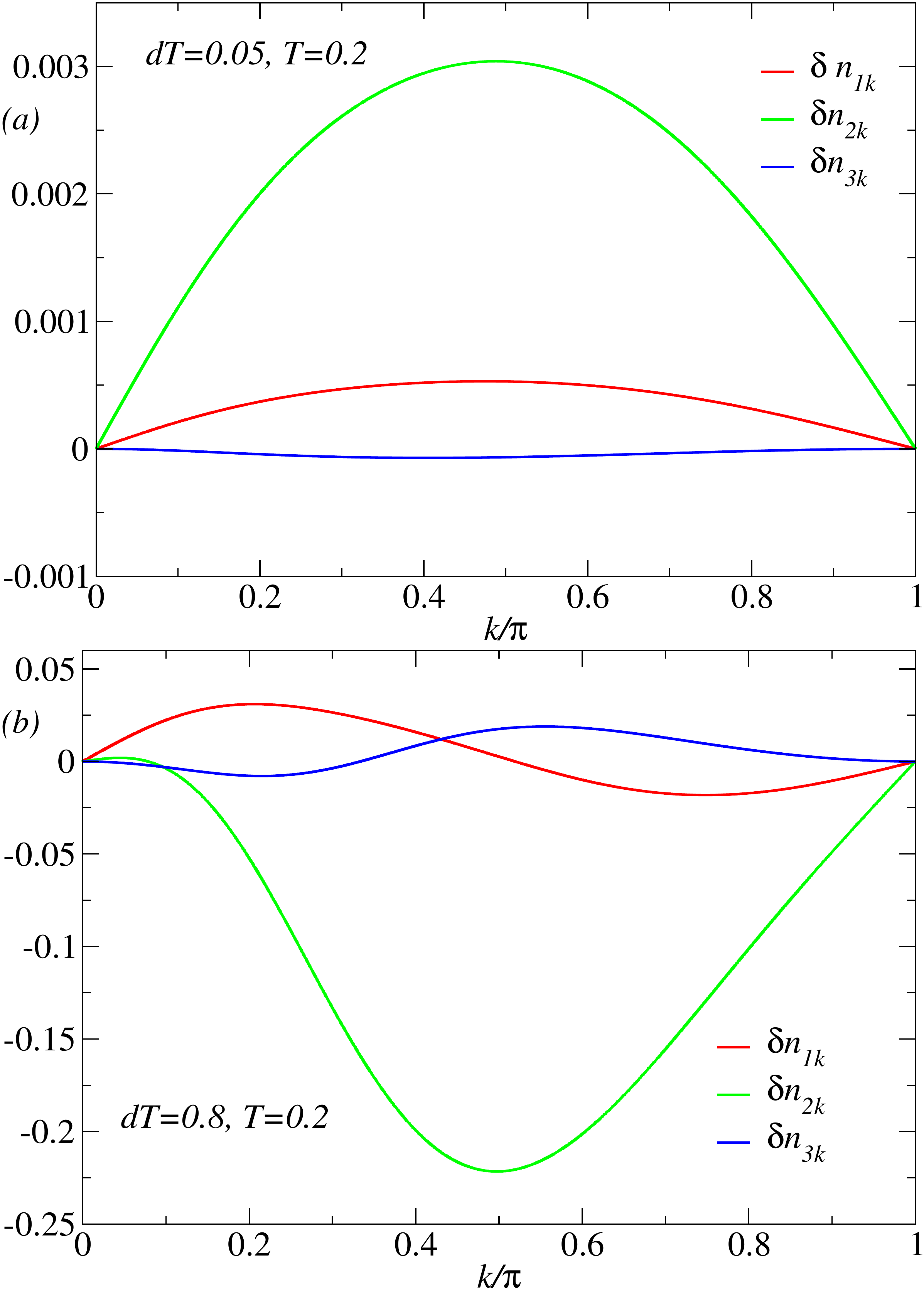}} 
\caption{$\de n_k=\ha_k(T,dT)-\ha_k(T,0)$ shown 
as a function of $k$. Here $\ha_k(T,dT)$ refers to the final value 
$\ha_\infty=\hb_\infty$ to which $\ha_m$ and $\hb_m$ settle down at large $m$, 
where the $\tau_i$'s in the unitary process are chosen according to the 
Thue-Morse sequence with given values of $T$ and $dT$. The drive parameters 
are (a) $g_i=4, g_f=2, T=0.2$ and $dT=0.05$, and (b) $g_i=4, g_f=2, T=0.2$ and 
$dT=0.8$. The components of $\de n_k$ are indicated as follows: $\de n_{1k}$ 
(red), $\de n_{2k}$ (green) and $\de n_{3k}$ (blue).} 
\label{fig9main} \end{figure}

Finally, we show in Fig.~\ref{fig9main} the components of $\de n_k=\ha_k(T,dT)
-\ha_k(T,0)$ versus $k$ when both $\ha_m$ and $\hb_m$ have settled down to the 
same value for a unitary process with the $\tau_i$'s chosen according to the 
Thue-Morse sequence for a given $T$ and $dT$. Since $\ha_k(T,0)$ represents 
the unit vector that 
corresponds to the unitary matrix $U_k(T)$ for the perfectly 
periodic case ($dT=0$) given in Eqs.~\eqref{uk2} and \eqref{uk}, we must have 
$\de n_k \to 0$ as $dT \to 0$. In Fig.~\ref{fig9main} (a), we show the 
variation of $\de n_k$ versus $k$ for $dT=0.05, ~T=0.2$, and the corresponding
results for $dT=0.8, ~T=0.2$ (thus, $dT > T$) in Fig.~\ref{fig9main} (b). We 
see that indeed $\ha_\infty (=\hb_\infty)$ equals (up to a sign) the unit 
vector in the direction of $\ha_1+\hb_1$. The nature of 
the variation with $k$ is rather different for $dT \ll T$ (Fig.~\ref{fig9main}
(a)) and $dT \gg T$ (Fig.~\ref{fig9main} (b)). Also, we note that $\de n_k$ 
equals $0$ for $k=0$ and $k=\pi$ as expected since $|a_3|=1$ and $a_1=a_2=0$ 
independently of the value of $dT$ at these two momenta.

Knowing the value of $\ha_\infty=\hb_\infty$ as a function of $k$ 
(Fig.~\ref{fig9main}) allows us to construct the local description of the 
final non-equilibrium steady state. The construction is completely analogous 
to the perfectly
periodic driven case ($dT=0$) where the relevant integrals of motion are now 
$\mathcal{J}_k$ (in Eq.~\eqref{conservedJk}) at each momentum instead 
of $\mathcal{I}_k$ (in Eq.~\eqref{conservedIk}). The density matrix of the 
g-GGE then equals 
\bea \rho_{\mathrm{g-GGE}} ~=~ \frac{1}{Z} \exp(-\sum_k \lam_k \mathcal{J}_k),
\label{gGGE1} \eea 
where the Lagrange multipliers $\lam_k$ are fixed by the condition
\bea \mathrm{Tr}[\rho_{\mathrm{g-GGE}}\mathcal{J}_k] ~=~ \langle \psi_k(n=0)|
\mathcal{J}_k |\psi_k(n=0) \rangle, \label{gGGE2} \eea
and the normalization constant $Z$ ensures that $\mathrm{Tr}[\rho_{\mathrm{
g-GGE}}]=1$. The final steady state value of any local operator can then be 
evaluated by using this ensemble.

\begin{figure}
{\includegraphics[width=\hsize]{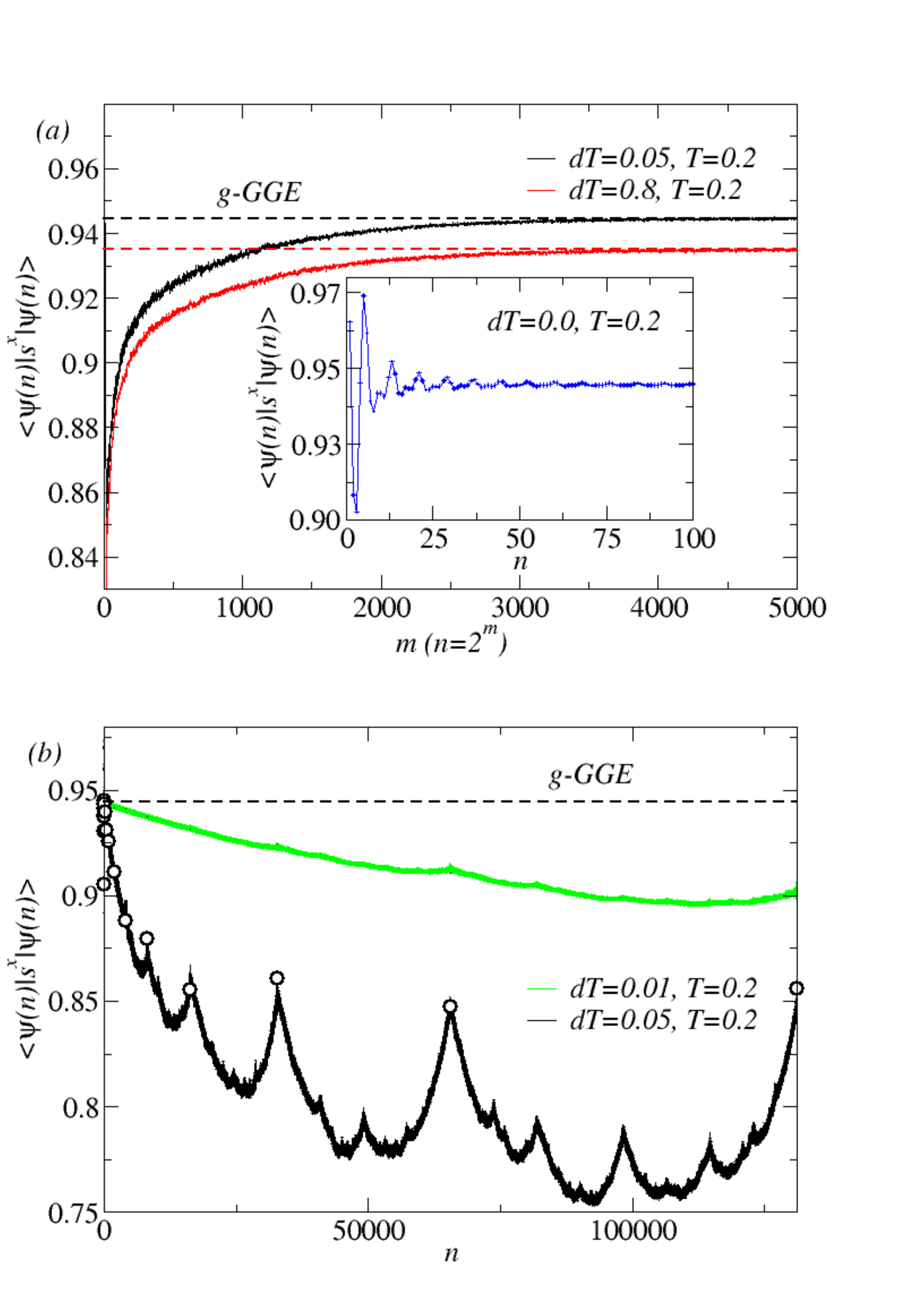}} 
\caption{(a) Convergence of $\langle 
\psi(n)|s^x|\psi(n) \rangle$ to the final steady state value calculated 
using the g-GGE. The steady state values are shown as dashed lines and are for
drive parameters $g_i=4,g_f=2, T=0.2$ and $dT=0.05$ (in black) and $dT=0.8$ 
(in red), with a system size equal to $L=131072$. Note 
that $m$ denotes exponentially growing time steps since $n=2^m$. The inset of 
(a) shows the rapid convergence of the same local quantity as a function of 
$n$ for a perfectly periodic protocol with $T=0.2$.
(b) The data for $dT=0.01$ (in green) and $dT=0.05$ (in black) 
now shown as a function of $n$ and the dashed line represents the 
value obtained from the g-GGE (the difference between the steady state values 
at $dT=0.01$ and $dT=0.05$ is small and not visible on this scale).
The open circles represent the 
calculation at $n=2^m$ using the unitary matrix $A_m$ at each $k$ 
for $dT=0.05$. The direct calculation using Eq.~\eqref{defsequence} 
at each $n$ (black line) agrees perfectly with the results at $n=2^m$ 
(open circles) using $A_m$.} \label{fig10main} \end{figure}

We show the results for one such local quantity 
$\langle \psi (n)|s^x |\psi(n) \rangle$ in Fig.~\ref{fig10main}.
Fig.~\ref{fig10main} (a) shows the convergence of this quantity 
to the final steady state value calculated using $\rho_{\mathrm{g-GGE}}$ 
for two different cases ($dT \ll T$ and $dT \gg T$) as a function of $n=2^m$; 
the convergence to the final g-GGE is extremely slow as we have already 
discussed in this section. Fig.~\ref{fig10main} (b) shows the 
data for the case $dT \ll T$ as a function of $n$ from which 
it is clear that even after a large number of drives $n \sim 10^5$, the 
system does not seem to relax to the final non-equilibrium steady state in 
any simple manner if $dT \neq 0$ (whereas the relaxation of 
local quantities at $dT=0$ is much faster as is clear from the inset of 
Fig.~\ref{fig10main} (a)), and the relaxation to the non-equilibrium steady 
state takes place over a much longer time scale, as we see in 
Fig.~\ref{fig10main} (a). 
We note here that this extremely slow relaxation of local quantities to 
their final steady state appears to be quite insensitive to the value of $dT$ 
as long as it is non-zero. Fig.~\ref{fig10main} (b) also shows the 
perfect agreement of the data calculated at $n=2^m$ using the unitary matrix 
$A_m$ at each $k$ to the results of the more direct calculation at each $n$ 
(using Eq.~\eqref{defsequence}) without using the recursive structure of the 
Thue-Morse sequence.

\section{Conclusions and outlook}
\label{secVI}

In this work, we have considered a prototypical integrable model, the 
one-dimensional transverse field Ising model, which is continually driven with 
a time-dependent Hamiltonian that results in a unitary dynamics. Since the 
model is translation invariant, the modes can be labeled by their momentum $k$.
For a subsystem whose size is much smaller than that of the total system 
(shown schematically in Fig.~\ref{fig0main}), the reduced density matrix 
of the subsystem is controlled by certain coarse-grained quantities in 
momentum space as defined in Eq.~\eqref{coarsegrain1}. Even though the 
corresponding quantities for a single momentum continually fluctuate as $n \to
\infty$ and do not have a well-defined large-$n$ limit, these coarse-grained 
quantities have a well-defined $n \to \infty$ limit if a non-equilibrium 
steady state exists. Furthermore, we define a probability distribution on the 
unit sphere at each average momentum $k_c$ by employing a pseudospin 
representation at each $k$ and using the large number of momenta lying in a 
small cell in momentum space centered around a momentum $k_c$ during the 
coarse-graining procedure. This probability distribution at each $k_c$ 
changes in an {\it irreversible manner} to the final $n \to \infty$ 
distribution as a function of $n$. At large $n$, this distribution generated 
from the points in a cell (centered at $k_c$) at that $n$ has 
the property stated in Eq.~\eqref{twoprob} that it equals the probability 
distribution generated from the motion of the pseudospin at a single momentum 
$k_c$ (thus, with no coarse-graining in momentum) for a large number of drives.

We first recapitulate the results known for the well-studied case in which
the system is driven perfectly periodically with a time period $T$. 
Since the model is completely integrable with an
infinite number of conserved quantities $\mathcal{I}_k$ (defined in
Eq.~\eqref{conservedIk}), the system does not locally heat up to infinite 
temperature, and instead this driving leads to the periodic generalized Gibbs
ensemble. We also calculate the behavior of the coarse-grained quantities in 
momentum space and the probability distributions on the unit sphere as a 
function of $k_c$ which immediately shows that the non-equilibrium steady 
state is not an infinite temperature ensemble. 

We then consider two new driving protocols both of which involve driving with 
two possible periods $T+dT$ and $T-dT$, such that the unitary time evolution 
operators corresponding to the two periods do not commute with each other; 
these driving protocols can be thought of as perturbations of the perfectly 
periodic driving if $dT \ll T$. In the first driving protocol, the successive
drive periods are randomly chosen to be $T+dT$ and $T-dT$ with probability 
1/2 each. In this case, we find that after a large number of drives the system 
eventually approaches an infinite temperature ensemble. 
Remarkably, this ensemble also has the
property that the probability distribution of a coarse-grained quantity 
averaged over many momenta in a cell is equal to the probability distribution 
of the same quantity generated by many drives for a single momentum.
The second driving protocol that we consider is one in which the time 
periods $T+dT$ and $T-dT$ are chosen according to the Thue-Morse sequence. 
This sequence is neither periodic nor random, but it has the property of
being scale-invariant if the number of drives is doubled. In this case, we 
find that the system approaches a non-equilibrium steady state with conserved
quantities $\mathcal{J}_k$ (defined in Eq.~\eqref{conservedJk}) which are 
smooth functions of $k$ if we view it after $n = 2^m$
drives rather than after $n$ drives. We call this steady state a geometric 
generalized Gibbs ensemble; to the best of our knowledge, this
is the first known example of such an ensemble. We note however that this
ensemble emerges only when $m$ is large (of the order of 1300 for one
particular set of parameters) which implies that the number of drives
$n$ required is astronomically large. 

An interesting problem for future studies may be to understand how the local 
properties relax in time to the final geometric generalized Gibbs ensemble 
that we have discussed here. Such a study would lead to a 
deeper understanding of the fixed point of a sequence of $SU(2)$ matrices
which is generated from two arbitrary initial matrices $A_0$ and $B_0$ 
(which do not commute with each other) by the recursion relations given in 
Eq.~\eqref{abn1}. The numerical results that we have presented in 
Sec.~\ref{secV} provide only a glimpse of this general problem which may 
possibly involve a very rich mathematical structure.

More generally, it will be interesting to understand the conditions which 
determine whether a continually driven integrable system locally heats up to 
an infinite temperature ensemble or not~\cite{ongoing}. For the latter 
case, it will be instructive to find other examples of {\it quasiperiodic} 
drive protocols~\cite{Sutherland1986}, apart from the one we have pointed out 
here, where well-defined non-equilibrium steady states 
may emerge without the restriction of a time-periodic drive.

\section*{Acknowledgments}
A.S. is grateful to Kedar Damle and Pushan Majumdar for useful discussions. 
The work of A.S. is partly supported through the Partner Group program 
between the Indian Association for the Cultivation of Science (Kolkata)
and the Max Planck Institute for the Physics of Complex Systems (Dresden). 
D.S. thanks Department of Science and Technology, India for Project 
No. SR/S2/JCB-44/2010 for financial support.

\end{document}